\documentclass{elsarticle}

\usepackage{amsmath,amssymb,amsfonts}
\usepackage{graphicx}
\usepackage{textcomp}
\usepackage{xcolor}
\def\BibTeX{{\rm B\kern-.05em{\sc i\kern-.025em b}\kern-.08em
    T\kern-.1667em\lower.7ex\hbox{E}\kern-.125emX}}

\usepackage{booktabs} 
\usepackage{balance}
\usepackage{comment}
\usepackage[noend,linesnumbered,algoruled]{algorithm2e}
\usepackage{array}
\usepackage{makecell}
\usepackage{varwidth}
\usepackage{enumitem}
\usepackage{url}
\usepackage{hyperref}
\usepackage{wrapfig}
\usepackage{listings}
\usepackage{subfigure}
\usepackage{caption}
\usepackage{graphicx}
\usepackage{subfloat}
\usepackage{cleveref}
\usepackage{csquotes}
\usepackage{xspace}
\usepackage{colortbl}
\definecolor{cgreen}{rgb}{0.25,0.50,0.0}
\definecolor{cblue}{rgb}{0.5,0.7,0.94}
\definecolor{cgray}{rgb}{0,0,0}
\usepackage{wrapfig}
\usepackage{framed}
\colorlet{shadecolor}{blue!20}
\usepackage{geometry}
\usepackage[normalem]{ulem}

\usepackage{hyperref}
\hypersetup{
  colorlinks   = true, 
  urlcolor     = blue, 
  linkcolor    = blue, 
  citecolor   = blue 
}

\newcolumntype{?}{!{\vrule width 1.25pt}}
\newcommand{\bundle}[1]{\sett{#1}}

\newcommand{\pacas}{{\em PACAS}}

\newcommand{\ignore}[1]{}
\newcommand{\mc}[1]{\mathcal{ #1}}
\newcommand{\Qry}[2]{{#1}^{#2}}

\newcommand{\nit}[1]{\textit{#1}}
\newcommand{\tp}[1]{{#1}}
\newcommand{\attr}[1]{{\sf #1}}
\newcommand{\xyl}{(X,Y,L)}
\newcommand{\xy}{(X,Y)}
\newcommand{\sch}[1]{\mc{#1}}

\newcommand{\sett}[1]{{\bf #1}}
\newcommand{\val}[1]{{\sl #1}}
\newcommand{\dgh}{{\sf DGH}\xspace}
\newcommand{\dghs}{{\sf DGH}s\xspace}
\newcommand{\vgh}{{\sf VGH}\xspace}
\newcommand{\vghs}{{\sf VGH}s\xspace}
\newcommand{\Dom}{{\sf Dom}\xspace}
\newcommand{\dom}{{\sf dom}\xspace}

\newcommand{\ra}{\rightarrow}
\newcommand{\base}{\nit{base}}
\newcommand{\lca}{\nit{lca}}
\newcommand{\gpenalty}{\nit{penalty}}
\newcommand{\errors}{\ensuremath{\mc{E}}}
\newcommand{\cells}{\ensuremath{C}}

\newcommand{\Null}{\val{null}}

\newcommand{\client}{\nit{CL}\xspace}
\newcommand{\service}{\nit{SP}\xspace}
\newcommand{\budget}{\ensuremath{\mc{B}}\xspace}

\newcommand{\palg}{\textit{SafePrice}\xspace}
\newcommand{\calg}{\textit{SafeClean}\xspace}
\newcommand{\qedsymbol}{\rule{1ex}{1ex}}

\newcommand{\tbf}{\textbf{\textcolor{red}{X}}\xspace}

\newtheorem{example}{Example}[section]
\newtheorem{theorem}{Theorem}[section]
\newtheorem{definition}{Definition}
\newtheorem{proposition}{Proposition}
\newcommand{\boxtheorem}{\hfill $\square$\vspace*{0.2cm}}

\newcommand{\red}[1]{{\color{red}#1}}
\newcommand{\blue}[1]{{\color{black}#1}}

\definecolor{extgreen}{rgb}{0.0,0.6,0.0}
\newcommand{\green}[1]{{\color{extgreen}#1}}

\newcommand{\fei}[1]{\noindent \red{Comment(Fei): #1}}
\newcommand{\yu}[1]{\noindent \green{Comment(Yu): #1}}

\newcommand{\eat}[1]{}
\newcommand{\eattr}[1]{\noindent \begin{shaded} #1 \end{shaded}}
\newcommand{\eatttr}[1]{}

\begin{document}

\begin{frontmatter}
\title{Privacy-Aware Data Cleaning-as-a-Service}

\author[add1]{Yu Huang}
\ead{huang223@mcmaster.ca}
\author[add2]{Mostafa Milani}
\ead{mostafa.milani@uwo.ca}
\author[add1]{Fei Chiang}
\ead{fchiang@mcmaster.ca}


\address[add1]{Dept. of Computing and Software, McMaster University, Hamilton, Canada}
\address[add2]{Dept. of Computer Science, Western University, London, Canada }

\begin{abstract}
\blue{Data cleaning is a pervasive problem for organizations as they try to reap value from their data.}  Recent advances in networking and cloud computing technology have fueled a new computing paradigm called \emph{Database-as-a-Service}, where data management tasks are outsourced to large service providers.  In this paper, we consider a \emph{Data Cleaning-as-a-Service} model that allows a client to interact with a data cleaning provider who hosts curated, and sensitive data.  \blue{We present \pacas: a \underline{P}rivacy-\underline{A}ware data \underline{C}leaning-\underline{A}s-a-\underline{S}ervice model that facilitates interaction between the parties with client query requests for data, and a service provider using a data pricing scheme that computes prices according to data sensitivity.  We propose new extensions to the model to define \emph{generalized data repairs} that obfuscate sensitive data to allow data sharing between the client and service provider.   We present a new semantic distance measure to quantify the utility of such repairs, and we re-define the notion of consistency in the presence of generalized values.  The PACAS model uses \xyl-anonymity that extends existing data publishing techniques to consider the semantics in the data while protecting sensitive values.}  \blue{Our evaluation over real data  show that PACAS safeguards semantically related sensitive values, and provides lower repair errors compared to existing privacy-aware cleaning techniques.}
\end{abstract}

\end{frontmatter}

\section{Introduction} \label{sec:introduction}

Data cleaning is a pervasive problem motivated by the fact that real data is rarely error free.  Organizations continue to be hindered by poor data quality as they wrangle with their data to extract value. Recent studies estimate that up to 80\% of the data analysis pipeline is consumed by data preparation tasks such as data cleaning~\cite{hbr}.  A wealth of data cleaning solutions have been proposed to reduce this effort: constraint based cleaning that use dependencies as a benchmark to repair data values such that the data and dependencies are consistent \cite{BFFR05}, statistical based cleaning, which propose updates to the data according to expected statistical distributions \cite{dasu}, and leveraging master data as a source of ground truth \cite{fan12}.  

Recent advances in networking and cloud infrastructure have motivated a new computing paradigm called \emph{Database-as-a-Service} that lowers the cost and increases access to a suite of data management services. A \emph{service provider} provides the necessary hardware and software platforms to support a variety of data management tasks to a \emph{client}. Companies such as Amazon, Microsoft, and Google each provide storage platforms, accelerated computational capacity, and advanced data analytics services.  However, the adoption of data cleaning services has been limited due to privacy restrictions that limit data sharing. Recent data cleaning efforts that  use a curated, master data source share a similar service model to provide high quality data cleaning services to a client with inconsistent data~\cite{fan12,fan09\ignore{,fan11}}.  However, these techniques largely assume the master data is widely available, without differentiating information sensitivity among the attribute values.

\begin{figure}[ht]
\centering
\includegraphics[width=4.75in]{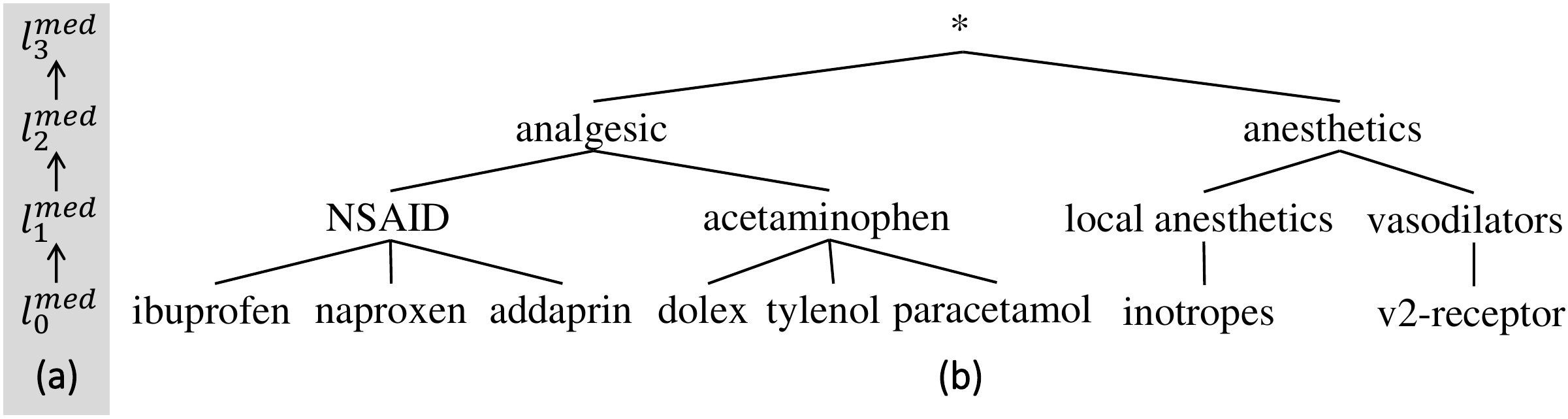}
  \caption{(a) $\dgh^\attr{med}$ and (b) $\vgh^\attr{med}$.}\label{fig:med} 
\end{figure}

\begin{figure}[ht]
\centering
  \includegraphics[width=7cm]{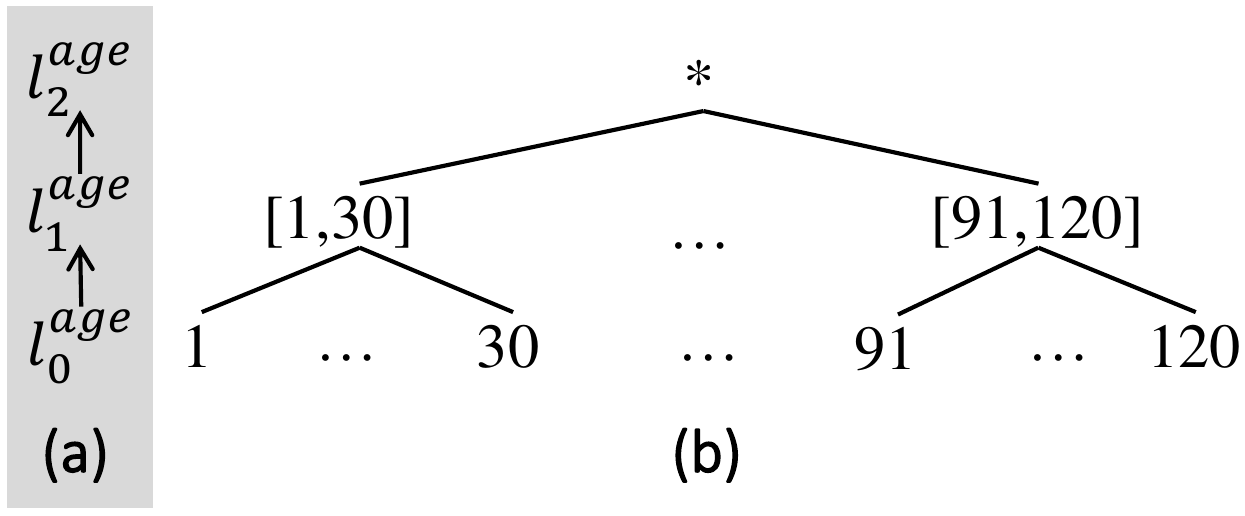}
  \caption{(a) $\dgh^\attr{age}$ and (b) $\vgh^\attr{age}$.}\label{fig:age}
\end{figure}

\begin{table}[ht]
\centering
\begin{tabular}{ | l | l | l | l | l |l|}
\hline
 \attr{ID}  &  \attr{GEN}    &  \attr{AGE} &  \attr{ZIP} &  \attr{DIAG}  &  \attr{MED}  \\
\hline \hline
$m_1$  &  male & 51 & P0T2T0 & osteoarthritis & ibuprofen \\
$m_2$  &  female & 45 & P2Y9L8 & tendinitis & addaprin  \\
$m_3$  &  female & 32 & P8R2S8 & migraine &  naproxen\\
$m_4$  &  female & 67 & V8D1S3 & ulcer &  tylenol  \\
  $m_5$  &  male & 61 & V1A4G1 & migraine &  dolex \\
$m_6$  &  male & 79 & V5H1K9 & osteoarthritis & ibuprofen \\
\hline
\end{tabular}
\vspace{-2mm}
\caption{\small Curated medical records ($R_{SP}$)}
\label{tab:master}
\vspace{4mm}
\begin{tabular}{ | l | l | l | l | l | l |}
\hline
 \attr{ID}  &  \attr{GEN}    &  \attr{AGE} &  \attr{DIAG}  &  \attr{MED}  \\
\hline \hline 
$t_1$  &      male \cellcolor{cgray!15} & 51 & osteoarthritis \cellcolor{cgray!15} & ibuprofen \cellcolor{cgray!15} \\
$t_2$  & male \cellcolor{cgray!15} & 79 &  osteoarthritis \cellcolor{cgray!15} & {\bf intropes} \cellcolor{cgray!15} \\
$t_3$  &      male \cellcolor{cgray!15}  & 45 & osteoarthritis \cellcolor{cgray!15}  & {\bf addaprin} \cellcolor{cgray!15}  \\
$t_4$  &      female \cellcolor{cgray!30} & 32 & {\bf ulcer} \cellcolor{cgray!30} & naproxen \cellcolor{cgray!30}  \\
$t_5$  &       female \cellcolor{cgray!30} & 67 & ulcer \cellcolor{cgray!30} &   tylenol \cellcolor{cgray!30} \\
$t_6$  &      male & 61 & migrane &  dolex  \\
$t_7$  &      female & 32 &  pylorospasm &  appaprtin \\
$t_8$  &       male & 37 & hypertension & dolex   \\
\hline
\end{tabular}
\vspace{-2mm}
\caption{\small Dirty client records w.r.t. $\varphi$.}
\label{tab:target}
\vspace{4mm}
\begin{tabular}{ | l | l | l | l | l |l|}
\hline
 \attr{ID}  &  \attr{GEN}    &  \attr{AGE} &  \attr{ZIP} &  \attr{DIAG}  &  \attr{MED}  \\
\hline \hline
$g_1$  & * & [31,60] & P* & osteoarthritis & ibuprofen \\
$g_2$  & * & [31,60] & P* & tendinitis & addaprin  \\
$g_3$  & * & [31,60] & P* & migraine & naproxen \\
$g_4$  & * & [61,90] & V* & ulcer &  tylenol  \\
$g_5$  & * & [61,90] & V* & migraine &  dolex \\
$g_6$  & * & [61,90] & V* & osteoarthritis & ibuprofen\\
\hline
\end{tabular}
\vspace{-2mm}
\caption{\small Public table.}
\label{tab:k}
\end{table}

\begin{example} \label{ex:intr} \em 
A data broker gathers longitudinal data from hospital and doctors' records, prescription and insurance claims. Aggregating and curating these disparate datasets lead to a valuable commodity that clients are willing to pay for gleaned insights, and to enrich and clean their individual databases.  Table \ref{tab:master} shows the curated data containing patient demographic, diagnosis and medication information. The schema consists of patient gender (\attr{GEN}), age (\attr{AGE}), zip code (\attr{ZIP}), diagnosed illness (\attr{DIAG}), and prescribed medication (\attr{MED}). 

A client such as a pharmaceutical firm, owns a stale and inconsistent subset of this data as shown in Table~\ref{tab:target}. For example, given \blue{a functional dependency (FD)} $\varphi:[\attr{GEN}, \attr{DIAG}]\rightarrow [\attr{MED}]$ on Table \ref{tab:target}, it states that a person's gender and diagnosed condition determine a prescribed medication.  \blue{That is, for any two tuples in  Table~\ref{tab:target}, if they share equal values in the [\attr{GEN}, \attr{DIAG}] attributes, then they should also have equal values in the \attr{MED} attribute.}  We see that tuples $\tp{t}_1$ - $\tp{t}_5$ falsify \blue{$\varphi$}.  \blue{Error cells that violate $\varphi$} are highlighted in light and dark gray, and values in bold are inaccurate according to Table~\ref{tab:master}.

If the client wishes to clean her data  with the help of a data cleaning service provider (i.e., the data broker and its data), she must first match the inconsistent records in Table \ref{tab:target} against the curated records in Table  \ref{tab:master}.  This generates possible fixes (also known as repairs) to the data in Table \ref{tab:target}. The preferred repair is to update $t_2[\attr{MED}]$, $t_3[\attr{MED}]$ to  \val{ibuprofen} (from $m_{6}$), and $t_4[\attr{DIAG}]=$ \val{migraine} (from $m_{3}$). However, this repair discloses sensitive patient information about diagnosed illness and medication from the service provider. \blue{It may be preferable to disclose a more general value that is semantically similar to the true value to protect individual privacy.  For example, instead of disclosing the medication  \val{ibuprofen}, the service provider discloses \val{Non-steroid anti-inflammatory drug (NSAID)}, which is the family of drugs containing \val{ibuprofen}.  In this paper, we explore how to compute such generalized repairs in an interactive model between a service provider and a client to protect sensitive data and to improve accuracy and consistency in client data.}


\end{example}


\textbf{State-of-the-Art.} Existing work in data privacy and data cleaning have been limited to imputation of missing values using decision trees~\cite{jagannathan}, or \blue{information theoretic} techniques~\cite{CG18}, and studying trade-offs between privacy bounds and query accuracy over differentially private relations~\cite{krishnan}.  In the \emph{data cleaning-as-a-service} setting, which we consider in this paper, using differential privacy poses the following limitations: (i) differential privacy provides guarantees assuming a limited number of interactions between the client and service provider; (ii) queries are limited to aggregation queries; and (iii) data randomization decreases data utility.

\eatttr{
\textbf{State-of-the-Art.} There has been limited work that considers data privacy requirements in data cleaning. Jaganathan and Wright propose a privacy-preserving protocol between two parties that imputes missing data using a lazy decision-tree imputation algorithm \cite{jagannathan}. PrivateClean provides data cleaning on local differentially private relations using a set of user-defined cleaning operations with trade-offs between privacy bounds and query accuracy~\cite{krishnan}.  In a \emph{data cleaning-as-a-service} setting, where a client and service provider interact over multiple iterations, using differential privacy poses the following limitations: (i) differential privacy provides guarantees assuming a one-time interaction between the two parties; (ii) queries are limited to aggregation type queries; and (iii) data randomization in differential privacy decreases data utility during data cleaning.\\}

Given the above limitations, we explore the use of \emph{Privacy Preserving Data Publishing} (PPDP) techniques, that even though do not provide the same provable privacy guarantees as differential privacy, do restrict the disclosure of sensitive values without limiting the types of queries nor the number of interactions between the client and service provider.  PPDP models prevent re-identification and break attribute linkages in a published dataset by hiding an individual record among a group of other individuals. This is done by removing identifiers and generalizing {\em quasi-identifier} (QI) attributes (e.g.,  \attr{GEN}, \attr{AGE}, \attr{ZIP}) that together can re-identify an individual. Well-known PPDP methods such as {\em $k$-anonymity} require the group size to be at least $k$ such that an individual cannot be identified from $k-1$ other individuals~\cite{sweeney,samarati}.  Extensions include {\em \xy-anonymity} that break the linkage between the set $X$ of QI attributes, and the set $Y$ of sensitive attributes by requiring at least $k$ distinct sensitive values for each unique $X$~\cite{fung}.  For example, Table~\ref{tab:k} is $k$-anonymous for $k=3$ by generalizing values in the QI attributes. It is also \xy-anonymous for sensitive attribute \attr{MED} for $k = 3$ since there are three distinct medications for each value in the QI attributes $X$, e.g., values $(*,[31,60],P*)$ of $X$ co-occur with \val{ibuprofen}, \val{addaprin}, and \val{naproxen} of $Y$.

To apply PPDP models in data cleaning, we must also address the highly contextualized nature of data cleaning, where domain expertise is often needed to interpret the data to achieve correct results.  It is vital to incorporate these domain semantics during the cleaning process, and into a privacy model during privacy preservation.  Unfortunately existing PPDP models only consider syntactic forms of privacy via generalization and suppression of values, largely ignoring the data semantics.  \blue{For example, upon closer inspection of Table \ref{tab:k}, the values in the QI attributes in records $g_{1} - g_{3}$ are associated with the same family of medication, since  \val{ibuprofen}, \val{addaprin}, and \val{naproxen} all belong to the the \val{NSAID} class, which are \val{analgesic} drugs.}  \blue{In past work, we introduced a privacy-preserving framework that uses an extension of \xy-anonymity to incorporate a \emph{generalization hierarchy}, which capture the semantics of an attribute domain}\footnote{$k$-anonymity and extensions, e.g.,  $l$-diversity and $t$-closeness also do not consider the underlying data semantics.}~\cite{huang2018pacas}.   In Table \ref{tab:k}, the medications in $g_{1} - g_{3}$ are modeled as synonyms in such a hierarchy.  \blue{In this paper, we extend our earlier work~\cite{huang2018pacas} to define a semantic distance metric between values in a generalization hierarchy, and to incorporate generalized values as part of the data repair process.}

\begin{example} \em 
To clean Table \ref{tab:target}, a client requests the correct value(s) in tuples $t_1 - t_3$ from the data cleaning service provider.  The service provider returns \val{ibuprofen} to the client to update $t_{2}[\attr{MED}]$ and $t_{3}[\attr{MED}]$ to $m_6[\attr{MED}]=$ \val{ibuprofen}.  However, disclosing the specific \val{ibuprofen} medication violates the requirements of the underlying privacy model. \blue{Hence, the service provider must determine when such cases arise, and offer a revised solution to disclose a less informative, generalized value, such as \val{NSAID} or \val{analgesic}.  The model must provide a mechanism for the client and service provider to negotiate such a trade-off between data utility and data privacy.}




\end{example}

\noindent  \textbf{Technical Challenges.}  
 \begin{enumerate}[nolistsep, leftmargin=*]
\item Data cleaning with functional dependencies (FDs), and record matching (between the service provider and client) is an NP-complete problem.  In addition, it is approximation-hard, i.e., the problem cannot be approximately solved with a polynomial-time algorithm and a constant approximation ratio~\cite{fan12}.  We extend this data cleaning problem with privacy restrictions on the service provider, making it as hard as the initial problem.  \blue{In our earlier work, we analyzed the complexity of this problem, and proposed a data cleaning framework realizable in practice~\cite{huang2018pacas}.  In this paper, we extend this framework to consider a new type of repair with generalized values that further protects sensitive data values.}

\item Given a generalization hierarchy as shown in Figures~\ref{fig:med} and \ref{fig:age}, proposed repairs from a service provider may contain general values that subsume a set of specific values at the leaf level.  In such cases, we study \blue{and propose} a new (repair) semantics for data consistency with respect to (w.r.t.) an FD involving general values.

\ignore{
The \xyl-anonymity model extends (X,Y)-anonymity as a privacy model in {\em privacy-preserving data publishing} (PPDP) where the goal is to publish a database that satisfies (X,Y)-anonymity or other privacy models and has maximum utility according to some utility measures~\cite{wang12}. 
We intend to use \xyl-anonymity in an interactive setting where the client decides about the data to be disclosed by asking a sequence of queries. In this new setting, we need to specify: (i) what it means to protect \xyl-anonymity, (ii) how a pricing scheme can protect \xyl-anonymity, and (iii) a pricing schemes that in fact guarantees this protection.
}

\item \blue{By proposing generalized values as repair values, we lose specificity in the client data instance.  We study the trade-off between generalized versus specific repair values. Resolving inconsistencies w.r.t. a set of FDs requires traversing the space of possible fixes (which may now include general values).  Quantifying the individual utility of a generalized value as a repair value, while respecting data cleaning budgets from the client is our third challenge.}

 \eat{Accessing a service provider for data repair reduces the space of fixes but it still lefts the repair algorithm with values to be asked and their of generalization according to the generalization hierarchies. A repair algorithm that applies generalized values as reliable fixes from the service provider comes at a loss of specificity. We study the trade-off between reliability (or confidence) of a value versus specificity. }


\end{enumerate}



\noindent \textbf{Our Approach and Contributions.} 
We build upon our earlier work that introduced \pacas, a $P$rivacy-$A$ware data $C$leaning-$A$s-a-$S$ervice framework that facilitates data cleaning between a client and a service provider~\cite{huang2018pacas}. The interaction is done via a data pricing scheme where the service provider charges the client for each disclosed value, according to its adherence to the privacy model.  PACAS includes a new privacy model that extends \xy-anonymity to consider the data semantics, while providing stronger privacy protection than existing PPDP methods.  \blue{In this work, we present extensions to \pacas\ to resolve errors (FD violations) by permitting updates to the data that are generalizations of the true value to avoid disclosing details about sensitive values.  Our goal is to provide repair recommendations that are semantically equivalent to the true value in the service provider data.}  We make the following contributions: 



\begin{enumerate}[nolistsep, leftmargin=*]
\item \blue{We extend \pacas \, the existing privacy-preserving, data cleaning framework that identifies \blue{FD errors} in client data, and allows the client to purchase clean, curated data from a service provider~\cite{huang2018pacas}.  We extend the repair semantics to introduce \emph{generalized repairs} that update values in the client instance to generalized values.}

\item \blue{We re-define the notion of consistency between a relational instance (with generalized values) and a set of FDs.  We propose an entropy-based measure that quantifies the semantic distance between two values to evaluate the utility of repair candidates.}



\item We \blue{extend} our \calg \ algorithm that resolves \blue{FD errors} by using external data purchased from a service provider.  \calg \ proposes repairs to the data by balancing its data privacy requirements against satisfying query purchase requests for its data at a computed price~\cite{huang2018pacas}.  Given a cleaning budget, we present a new budget allocation algorithm that improves upon previous, fixed allocations, to consider allocations according to the number of errors in which a database value participates. 



\item \blue{We evaluate the effectiveness of \palg and \calg over real data showing that by controlling data disclosure via data pricing, we can effectively repair FD violations while guaranteeing \xyl-anonymity.  We also show that our \calg achieves lower repair error than comparative baseline techniques. 
}
\end{enumerate}

We provide notation and preliminaries in Section \ref{sec:background}.  \blue{In Section~\ref{sec:definitions}, we define generalized relations, and their consistency w.r.t. a set of FDs, and present a new semantic distance function.} We present our problem definition, and review the \pacas \ system in Section~\ref{sec:pacas}. \blue{We discuss how data privacy is enforced via data pricing in Section~\ref{sec:disclosure}, and describe new extensions to our data cleaning algorithm that consider generalized values} in Section~\ref{sec:framework}. We present our experimental results in Section~\ref{sec:experiments}, related work in Section~\ref{sec:related}, and conclude in Section~\ref{sec:conc}.
\section{Preliminaries} \label{sec:background}

\ignore{\begin{figure*}
\begin{minipage}{.65\textwidth}
  \begin{center}
\hspace{-4mm}\includegraphics[width=5.0in]{./fig/diag}
\vspace{-5mm}
  \caption{(a) $\dgh^\attr{diag}$ and (b) $\vgh^\attr{diag}$.}\label{fig:diag} 
  \end{center}
\end{minipage}
\hspace{4mm}
\begin{minipage}{.33\textwidth}
    \begin{center}
  \vspace*{13mm}
  \includegraphics[width=3.25cm]{./fig/gen}
  \vspace{-1mm}
  \caption{(a) $\dgh^\attr{gen}$ and (b) $\vgh^\attr{gen}$.}\label{fig:gen}
  \end{center}
\end{minipage}
\end{figure*}}


\subsection{Relations and Dependencies}
\label{sec:depdefn}
A {\em relation} (table) $R$ with a {\em schema} $\sch{R}=\{A_1,...,A_n\}$ is a finite set of $n$-ary tuples $\{\tp{t}_1,...,\tp{t}_N\}$. A {\em database} ({\em instance}) $D$ is a finite set of relations $R_1,...,R_m$ with schema $\blue{\mathfrak{R}}=\{\sch{R}_1,...\sch{R}_m\}$. We denote by small letters $x,y,z$ as variables. Let $A,B,C$ refer to single attributes and $X,Y,Z$ as sets of attributes. A {\em cell} $c=t[A_i]$ is the $i$-th position in tuple $\tp{t}$ with its value denoted by $c.{\sf value}$. We use $c$ to refer to the value $c.{\sf value}$ if it is clear from the context. A {\em functional dependency} (FD) $\varphi$ over a relation $R$ with set of attributes $\sch{R}$ is denoted by $\varphi: X \rightarrow Y$, in which $X$ and $Y$ are subsets of $\sch{R}$. We say $\varphi$ holds over $R$, $R\models \varphi$, if for every pair of tuples $\tp{t}_1$, $\tp{t}_2$ $\in R$, $\tp{t}_1[X]=\tp{t}_2[X]$ implies $\tp{t}_1[Y]=\tp{t}_2[Y]$. Table~\ref{tab:symbols} summarizes our symbols and notations.

A {\em matching dependency} (MD) $\phi$ over two relations $R$ and $R'$ with schemata $\mc{R}=\{A_1,A_2,...\}$ and $\mc{R}'=\{A'_1,A'_2,...\}$ is an expression of the following form:

\vspace{-3mm}
\begin{align}
\bigwedge_{i \in [1,n]}R[X_i] \approx R'[X'_i] \rightarrow R[Y] \leftrightharpoons R'[Y'],\label{r:md}
\end{align}
\vspace{-2mm}

\noindent where $(X_i,X'_i)$ and $(Y,Y')$ are comparable pairs of attributes in $R$ and $R'$. The MD $\phi$ states that for a pair of tuples $(t,t')$ with $t\in R$ and $t' \in R'$, if $t[X'_i]$ values are similar to values $t'[X'_i]$ according to the similarity function $\approx$, the values of $t[Y]$ and $t'[Y']$ are identical~\cite{fan09}.

\begin{table}[]
  \centering
  \caption{Summary of notation and symbols.}
  \label{tab:symbols}
  \setlength\tabcolsep{2 pt}
  \small \begin{tabular}{l l}
  \toprule
    Symbol              & Description \\
    \midrule
    $R,\mc{R}$ & relation and relational schema\\
    $A,B$ & relational attributes\\
    $X,Y,Z$ & sets of relational attributes\\
    $D,\blue{\mathfrak{R}}$ & database instance and database schema\\    
    $\varphi,\phi$ & functional dependency and matching dependency\\    
    $\Dom^A,\dom^A$ & domain of attribute $A$\\
    $\dom^A(l)$ & sub-domain of attribute $A$ in level $l$\\
    $\dgh^A$,$\vgh^A$ & domain and value generalization hierarchies\\
    $\le$ & generalization relation for levels\\
    \blue{$\preceq$} & \blue{generalization relation for values, tuples, and relations}\\
    $Q,G$ & Simple query and generalized query (GQ)\\
    $l,L$ & level and sequence of levels\\
    $\mc{S},\mc{C}^Q$ & support set, and conflict set\\
    $\mc{B},B_i$ & total and the budget for the i-th iteration\\
    $l_\attr{max}$ & generalization level\\
    $c,e$ & database cell and error cell \ignore{\fei{i?}} \\  
    \blue{$\delta$} & \blue{distance function for values, tuples and relations}\\
    \bottomrule
    \end{tabular}
\end{table}

\subsection{Generalization} \label{sec:generalization}

Generalization replaces values in a private table with less specific, but semantically consistent values according to a generalization hierarchy. To generalize attribute $A$, we assume a set of levels $\mc{L}^A=\{l^A_0,...,l^A_h\}$ and a partial order $\le^A$, called a {\em generalization relation} on $\mc{L}^A$. Levels $l^A_i$ are assigned with disjoint domain-sets $\dom(l^A_i)$. In $\le^A$, each level has at most one parent. The domain-set $\dom(l^A_n)$ is the maximal domain set and it is a singleton, and $\dom(l^A_0)$ is the {\em ground domain set}. The definition of $\le^A$ implies the existence of a totally ordered hierarchy, called the {\em domain generalization hierarchy}, $\dgh^A$. The domain set $\dom(l^A_i)$ generalizes $\dom(l^A_j)$ iff $l^A_j \leq l^A_i$. We use $h^A$ to refer to the number of levels in $\dgh^A$\ignore{, i.e. $h^A=|\mc{L}^A|$}. Figures~\ref{fig:med}(a) and \ref{fig:age}(a) show the \dgh s for the medication and age attributes, resp. A {\em value generalization relationship} for $A$, is a partial order $\preceq^A$ on $\Dom^A=\bigcup\dom(l^A_i)$. It specifies a {\em value generalization hierarchy}, $\vgh^A$, that is a tree whose leaves are values of the ground domain-set $\dom(l^A_0)$ and whose root is the single value in the maximal domain-set $\dom(l^A_n)$ in $\dgh^A$. For two values $v$ and $v'$ in $\Dom^A$, $v' \preceq^A v$ means $v'$ is more specific than $v$ according to the \vgh. We use $\preceq$ rather than $\preceq^A$ when the attribute is clear from the context. The \vgh \ for the \attr{MED} and \attr{AGE} attributes are shown in Figures ~\ref{fig:med}(b) and \ref{fig:age}(b), respectively. According to \vgh of \attr{MED}, \val{ibuprofen} $\preceq$ \val{NSAID}.

A value is {\em ground} if there is no value more specific than it, and it is {\em general} if it is not ground. In Figure~\ref{fig:med}, \val{ibuprofen} is ground and \val{analgesic} is general. For a value $v$, its base denoted by $\attr{base}(v)$ is the set of ground values $u$ such that $u \preceq v$. We use $0 \le \attr{level}(v) \le h^A$ to refer to the level of $v$ according to $\vgh^A$.

A {\em general relation} (table) is a relation with some general values and a ground relation has only ground values. A {\em general database} is a database with some general relations and a ground database has only ground relations. The generalization relation $\preceq^A$ trivially extends to tuples. We give an extension of the generalization relation to general relations and databases in Section~\ref{sec:definitions}.

The generalization hierarchies (\dgh and \vgh) are either created by the data owners with help from domain experts or generated automatically based on data characteristics~\cite{kamber}. The automatic generation of hierarchies for categorical attributes~\cite{lee} and numerical attributes~\cite{fayyad,kerber,liuf} apply techniques such as histogram construction, binning, numeric clustering, and entropy-based discretization~\cite{kamber}.




\blue{\subsection{Generalized Queries}}\label{sec:gqueries}
We review {\em generalized queries} (GQs) to access values at different levels of the \dgh~\cite{huang2018pacas}.



\begin{definition}[Generalized Queries] \label{df:gq}\em A generalized query (GQ) with schema $\mc{R}$ is a pair $G=\langle Q,L\rangle$, where $Q$ is an $n$-ary select-projection-join query over $\mc{R}$, and $L=\{l_1,...,l_n\}$  is a set of levels for each of the $n$ values in $Q$ according to the \dghs in $\sch{R}$. The set of answers to $G$ over $R$, denoted as $G(R)$, contains $n$-ary tuples $\tp{t}$ with values at levels in $l_i \in \attr{L}$, such that  $\exists \tp{t}' \in Q(R)$ and $\tp{t}$ generalizes $\tp{t}'$, i.e. $\tp{t}' \preceq \tp{t}$.
\end{definition}


Intuitively, answering a GQ involves finding the answers of $Q(R)$, and then generalizing these values to levels that match $L$. For a fixed size \dgh, the complexity of answering $G$ is the same as answering $Q$.

\begin{example} \label{ex:gc} \em \blue{Consider a GQ $G=\langle Q,L\rangle$ with level $L=\{l_0^\attr{GEN}$, $l_1^\attr{MED}\}$ and query $Q(R_{SP})$ $=$ $\Pi_\attr{GEN,MED}(\sigma_{\attr{DIAG}=\val{migraine}}(R_{SP}))$ posed on relation $R_{SP}$ is Table~\ref{tab:master}. The query $Q$ requests the gender and medication of patients with a migraine. The answers to $Q(R_{SP})$ are \{(\val{female}, \val{naproxen}), (\val{male}, \val{dolex})\}. The GQ $G$ asks for the same answers but at the levels specified by $L$. Therefore the answers to $G$ are \{(\val{female}, \val{NSAID}),  (\val{male}, \val{acetaminophen})\}, which are generalized according to $L$ and Figure \ref{fig:med}.}
\end{example}  




\subsection{Privacy-Preserving Data Publishing}

The most well-known PPDP privacy model is $k$-anonymity that prevents re-identification of a single individual in an anonymized data set~\cite{sweeney,samarati}.

\begin{definition}[$X$-group and $k$-anonymity] \label{df:k} A relation $R$ is $k$-anonymous if every QI-group has at least $k$ tuples. An $X$-group is a set of tuples with the same values in $X$.\end{definition}

As an example, Table~\ref{tab:k} has two QI-groups, $\{g_1,g_2,g_3\}$ and $\{g_4,g_5,g_6\}$, and it is $k$-anonymous with $k=3$.  $k$-anonymity is known to be prone to attribute linkage attacks where an adversary can infer sensitive values given QI values.  Techniques such as \xy-anonymity aim to address this weakness by ensuring that for each QI value in $X$, there are at least $k$ different values of sensitive values in attribute(s) $Y$ in the published (public) data~\blue{\cite{seq-wang}}.  



\begin{definition} [\xy-anonymity] \em \label{df:xy} A table $R$ with schema $\sch{R}$ and attributes $X,Y \subseteq \sch{R}$ is {\em \xy-anonymous} with value $k$ if for every $\tp{t} \in R$, there are at least $k$ values in $\Qry{Q}{t}(R)$, $|\Qry{Q}{t}(R)| \ge k$, where $\Qry{Q}{t}(R)=\Pi_{Y}(\sigma_{X=t[X]}(R))$. \end{definition}

\blue{The query $Q^t(R)$ in Definition~\ref{df:xy} returns distinct values of attributes $Y$ that appear in $R$ with the values $t[X]$. Therefore, if the size of $Q^t(R)$ is greater than $k$ for every tuple $t$, that means every values of $X$ are linked with at least $k$ values of $Y$. Note that $k$-anonymity is a special case of \xy-anonyity when $X$ is the set of QI attributes and $Y$ is the set of sensitive attributes that are also a key in $R$~\cite{seq-wang}.}

\begin{example}\em \label{ex:xyweakness}
For $X=\{\attr{GEN}, \attr{AGE}, \attr{ZIP}\}$, $Y=\{\attr{MED}\}$ with $k=3$, Table \ref{tab:k} is \xy-anonymous since each $g_i \in R$, $\Qry{Q}{g_i}(R)$ is either \{\val{ibuprofen}, \val{addaprin}, \val{naproxen}\} or \{\val{tylenol}, \val{dolex}, \val{ibuprofen}\}\blue{, which means the values of $X$ in each tuple $g_i$ are linked to $k=3$ distinct values of $Y$ ($Q^{g_i}(R)$ represents the set of distinct values of $Y$ that are linked to the values of $g_i[X]$ in $R$).}  Table \ref{tab:k} is not \xy-anonymous with $X=\{\attr{DIAG}\}$ and $Y=\{\attr{MED}\}$, since for $g_1$, $\Qry{Q}{g_1}(R)=\{$\val{ibuprofen}$\}$ with size $1 \le k$.
\end{example}



The \xy-anonymity model extends $k$-anonymity; if $Y$ is a key for $R$ and $X,Y$ are QI and sensitive attributes, respectively, then \xy-anonymity reduces to $k$-anonymity.

The $l$-diversity privacy model extends $k$-anonymity with a stronger restriction over the $X$-groups~\cite{mach}.  A relation is considered $l$-diverse if each $X$-group contains at least $l$ ``well-represented" values in the sensitive attributes $Y$.  Well-representation is normally defined according to the application semantics, e.g., {\em entropy $l$-diversity} requires the entropy of sensitive values in each  $X$-group to satisfy a given threshold~\cite{mach}.  When well-representation requires $l$ sensitive values in each $Y$ attribute, \xy-anonymity reduces to $l$-diversity.



\noindent \blue{\subsubsection{\xyl-anonymity.}}
\label{sec:xyl}
In earlier work, we introduced \xyl-anonymity, which extends \xy-anonymity to consider the semantic closeness of values~\cite{huang2018pacas}.  \blue{Consider the following example showing the limitations of \xy-anonymity.}

\blue{\begin{example}\label{ex:xy-issue} \xy-anonymity prevents the linkage between sets of attributes $X$ and $Y$ in a relation $R$ by making sure that every values of $Y$ appears with at least $k$ distinct values of $X$ in $R$. For example, Table~\ref{tab:k} is \xy-anonymous with $X=\{\attr{GEN},\attr{AGE},\attr{ZIP}\}$, $Y=\{\attr{MED}\}$, and $k=3$ because the values of $X$ appear with three different values; $(*,[31,60],P*)$ co-occurs with \val{ibuprofen}, \val{addaprin}, \val{naproxen} and $(*,[61,90],V*)$ appears with \val{tylenol}, \val{dolex}, \val{ibuprofen}. 

\xy-anonymity ignores how close the values of $Y$ are according to their \vghs. For example, although \val{ibuprofen}, \val{addaprin}, \val{naproxen} are distinct values, they are similar medications with the same parent \val{NSAID} according to the \vgh in Figure~\ref{fig:med}. This means the value $(*,[31,60],P*)$ of attributes $X$ is linked to \val{NSAID} which defies the purpose of \xy-anonymity.

\xyl-anonymity resolves this issue by adding a new parameter $L$ for levels of $Y$. In \xyl-anonymity, every values of $X$ has to appear with at least $k$ values of $Y$ in the levels of $L$. For example, Table~\ref{tab:k} is not \xyl-anonymous if $L=\{l_1^\attr{MED}\}$ because $(*,[31,60],P*)$ appears with $Y$ values that all roll up to \val{NSAID} at level $l_1^\attr{MED}$. Thus, to achieve \xyl-anonymity we need to further suppress values in Table~\ref{tab:k} to prevent the linkage between $(*,[31,60],P*)$ and \val{NSAID}.\end{example}}

\begin{definition}[\xyl-anonymity] \label{df:xyl} \em Consider a table $R$ with schema $\sch{R}$ and attributes $X,Y \subseteq \sch{R}$, and a set of levels $L$ corresponding to attribute \dghs from $Y$.  $R$ is {\em \xyl-anonymous} with value $k$ if for every $t \in R$, there are at least $k$ values in $\Qry{G}{t}(R)$, where $\Qry{G}{t}=\langle \Qry{Q}{t},L\rangle$ is a GQ with $\Qry{Q}{t}(R)=\Pi_{Y}(\sigma_{X=t[X]}(R))$. \end{definition}

\blue{In Definition~\ref{df:xyl}, the GQ $G^t(R)$ returns the set of $Y$ values in levels $L$ that appear with the $X$ values of the tuples $t$.
If the size of $G^t(R)$ is greater than $k$, then every value of $X$ appears with at least $k$ values of $Y$ in the level $L$, which is the objective of \xyl-anonymity, as shown in Example~\ref{ex:xy-issue}.}
\ignore{Intuitively, $R$ is \xyl-anonymous if tuples in each $X$-group have at least $k$ different values of $Y$ at level $L$.}


\ignore{\begin{figure}
 \begin{center}
\includegraphics[width=\linewidth]{./fig/med.jpg}
\vspace{-4mm}
  \caption{\dgh and \vgh for medications.}\label{fig:med}
\vspace{-4mm}
\end{center}
\end{figure}}

\subsection{Data Pricing} \label{sec:pricingpr}

High quality data is a valuable commodity that has lead to increased purchasing and selling of data online.  \blue{Data market services provide data across different domains such as geographic and locational (e.g., AggData~\cite{aggdata}), advertising (e.g., Oracle~\cite{oracle}, Acxiom~\cite{acxiom}), social networking (e.g., Facebook~\cite{facebook}, Twitter~\cite{twitter}), and people search (e.g. Spokeo~\cite{spokeo}).  These vendors have become} popular in recent years, and query pricing has been proposed as a fine-grained and user-centric technique to support the exchange and marketing of data~\cite{balazinska}.


Given a database instance $D$ and query $Q$\ignore{(or $\bundle{Q}$)}, a {\em pricing function} returns a non-negative real number representing the price to answer $Q$\ignore{ (or $\bundle{Q}$)}~\cite{balazinska}.  A pricing function should have the desirable property of being {\em arbitrage-free}~\cite{balazinska}.  Arbitrage is defined using the concept of {\em query determinacy}~\cite{nash}. Intuitively, $Q$ determines $Q'$ if for every database $D$, the answers to $Q'$ over $D$ can be computed from the answers to $Q$ over $D$. Arbitrage occurs when the price of $Q$ is less than that of $Q'$, which means someone interested in purchasing $Q'$ can purchase the cheaper $Q$ instead, and compute the answer to $Q'$ from $Q$. For example, $Q(R)=\Pi_{Y}(R)$ determines $Q'(R)=\Pi_{Y}(\sigma_{Y < 10}(R))$ because the user can apply $Y<10$ over $Q(R)$ to obtain $Q'(R)$. Arbitrage occurs if $Q$ is cheaper than $Q'$ which means a user looking for $Q'(R)$ can buy $Q$ and compute answers to $Q'$. An arbitrage-free pricing function denies any form of arbitrage and returns consistent pricing without inadvertent data leakage~\cite{balazinska}. A pricing scheme should also be {\em history-aware}~\cite{balazinska}.  A user should not be charged multiple times for the same data purchased through different queries. In such a pricing scheme, the data seller must track the history of queries purchased by each buyer and price future queries according to historical data.  
\section{Generalized Relations} \label{sec:definitions}
Using general values as repair values in data cleaning requires us to re-consider the definition of consistency between a relational instance and a set of FDs.  In this section, we introduce an entropy-based measure that allows us to quantify the utility of a generalized value as a repair value, and present a revised definition of consistency that has not been considered in past work~\cite{huang2018pacas}.


\eat{We extend the generalized hierarchy idea in \cite{huang2018pacas} and introduce an entropy-based metrics to measure the semantic distance between two generalized values at different hierarchy levels. We also present a revised definition of consistency between a generalized relation and a set of FDs. }

\eat{In this section, we give some new definition related to generalized relation that are needed for our problem definition in Section~\ref{sec:pacas}. We first present a semantic measure of distance between general values in Section~\ref{sec:dist} and then we define consistency of a general relation w.r.t. FDs in Section~\ref{sec:GDBcons}.}


\subsection{Measuring Semantic Distance} \label{sec:dist}
By replacing a value $v'$ in a relation $R$ with a \emph{generalized} value $v$, there is necessarily some information loss.  We present an entropy-based penalty function that quantifies this loss~\cite{gionis}.  We then use this measure as a basis to define a distance function between two general values. 


\blue{\begin{definition}[Entropy-based Penalty~\cite{gionis}]\label{df:penalty}
Consider an attribute $A$ in a ground relation $R$. Let $X_A$ be a random variable to randomly select a value from attribute $A$ in $R$.  Let $\vgh^A$ be the value generalization hierarchy of the attribute $A$ (the values of $A$ in $R$ are ground values from $\vgh^A$). The entropy-based penalty of a value $v$ in $\vgh^A$ denoted by $E(v)$ is defined as follows: 

$$E(v)=P(X_A \in \base(v)) \times H(X_A|X_A \in \base(v)),$$
\noindent where $P(X_A \in \base(v))$ is the probability that the value of $X_A$ is a ground value and a descendant of $v$, $X_A \in \base(v))$. The value $H(X_A|X_A \in \base(v))$ is the entropy of $X_A$ conditional to $X_A \in \base(v)$.\ignore{ and is defined as follows~\cite{gionis}:

$$H(X_A|X_A \in \base(v))=-\sum_{a_i\in R[A]}P_{a_i}\times \log P_{a_i}.$$

\noindent Here, $P_{a_i}=P(X_A=a_i|X_A \in \base(v))$ is the probability of $P(X_A=a_i)$ conditional to $X_A \in \base(v)$.}
\end{definition}}




Intuitively, \blue{the entropy $H(X_A|X_A \in \base(v))$} measures the uncertainty of using the general value $v$. $E(v)$ measures the information loss of replacing values in $\base(v)$ with $v$. Note that $E(v)=0$ if $v$ is ground \blue{because $H(X_A|X_A \in \base(v))=0$,} and $E(v)$ is maximum if $v$ is the root value $*$ in $\vgh^A$. $E(v)$ is monotonic whereby if $v\preceq v'$, then $E(v) \le E(v')$. Note that the conditional entropy \blue{$H(X_A|X_A \in \base(v))$} is not a monotonic measure~\cite{gionis}. \ignore{The penalty for a generalized table $R$, i.e. $\gpenalty(R)$, is the sum of $E(v)$ for every $v \in R$, where $\gpenalty(R) = 0$ if $R$ is a ground relation, and is greater than zero if it is generalized.} 

\blue{\begin{example}\em \label{ex:penalty}Consider the general value $v = g_1[\attr{AGE}]=$[31,60] in Table~\ref{tab:master}. Three ground values \val{51}, \val{45} and \val{32} in  $base(v)$ appear in Table~\ref{tab:master}, which has total $6$ records, so $P(X_\attr{AGE} \in base(v)) = \frac{3}{6} = \frac{1}{2}$. According to Table~\ref{tab:master}, the conditional entropy is $H(X_A|X_\attr{AGE} \in base(v))=3\times (-\frac{1}{3}\times \log \frac{1}{3})= 1.58$ because \val{51}, \val{45} and \val{32} each appear exactly once in the table. Therefore, the entropy-based penalty of $v$ is $E(v)=\frac{1}{2}\times1.58=0.79$.  
\end{example}}


\blue{\begin{definition}[Semantic Distance Function, $\delta$] \label{df:distance} The semantic distance between $v$ and $v'$ in $\vgh^A$ is defined as follows:

\vspace{-4mm}
\begin{align*}
\delta(v,v')=
    \begin{cases}
    |E(v') - E(v)| &\textit{if}\; v\preceq v'\;\textit{or}\; v'\preceq v\\
    \delta(v,a)+\delta(a,v') &\textit{otherwise}
    \end{cases}
\end{align*}
 
\noindent in which $a=\lca(v,v')$ is the least common ancestor of $v$ and $v'$.
\end{definition}}

\blue{Intuitively, if $v$ is a descendant of $v'$ or vice versa, i.e. $v\preceq v'$ or $v'\preceq v$, their distance is the the difference between their entropy-based penalties, i.e., $|E(v') - E(v)|$. This is the information loss incurred by replacing a more informative child value $v$ with its ancestor $v'$ when $v \preceq v'$. If $v$ and $v'$ do not align along the same branch in the \vgh, i.e. $v\not\preceq v'$ and $v'\not\preceq v$, $\delta(v,v')$ is the total distance between $v$ and $v'$ as we travel through their least common ancestor $a$, i.e. $\delta(v,a)+\delta(a,v')$.}  



\blue{\begin{example}\em \label{ex:distance} (Ex.~\ref{ex:penalty} continued) According to Definition~\ref{df:distance}, $\delta([31,60], 51)=|E([31,60])-E(51)|=|0.79-0|=0.79$ because $51 \preceq [31,60]$. Similarly, $\delta([31,60], 45)=0.79$. Also, $\delta(45, 51)=\delta(45, [31,60])+\delta([31,60],51)=1.58$ because $45$ and $51$ do not belong to the same branch, and $[31,60]$ is their least common ancestor.  
\end{example}}



The $\delta(v'v)$ distance captures the semantic closeness between values in the \vgh. We extend the definition of $\delta$ to tuples by summing the distances between corresponding values in the two tuples. The $\delta$  function naturally extends to sets of tuples and relations. \blue{We use the $\delta(v'v)$ distance measure to define repair error in our evaluation in Section~\ref{sec:experiments}.}

\subsection{Consistency in Generalized Relations} \label{sec:GDBcons} 
\eat{We proposed a ground repair definition in \cite{huang2018pacas}. In this paper, we introduce a new definition of consistency w.r.t FDs in a generalized relation. }
A (generalized) relation $R$ may contain generalized values that are syntactically equal \blue{but are semantically different and represent different ground values.}  For example, $g_1[\attr{AGE}]$ and $g_2[\attr{AGE}]$ in Table~\ref{tab:k} are syntactically equal (containing [31,60]), but their true values in Table~\ref{tab:master} are different, $m_1[\attr{AGE}]=51$ and $m_2[\attr{AGE}]=45$, respectively. \blue{In contrast, two syntactically different general values may represent the same ground value. For example, although \val{analgesic} and \val{NSAID} are different general values, they may both represent \val{ibuprofen}.} The consistency between traditional FDs and a relation $R$ is based on syntactic equality between values. We present a revised definition of consistency with the presence of generalized values in $R$.


Given a generalized relation $R'$ with schema $\sch{R}$, and FD $\varphi: A \rightarrow B$ with attributes $A,B$ in $\sch{R}$, $R'$ satisfies $\varphi$ ($R' \models \varphi$), if for every pair of tuples $t_1,t_2$ in $R'$, if \blue{$t_1[A]$ and $t_2[A]$ are equal ground values then $t_1[B]$ is an ancestor of $t_2[B]$ or vice-versa, i.e. $t_1[B]\preceq t_2[B]$ or $t_2[B]\preceq t_1[B]$}.  Since a general value $v$ encapsulates a set of distinct ground values in $\base(v)$, relying on syntactic equality between two (general) values, $t_1[B]$ and $t_2[B]$ is not required to determine consistency, \blue{since they may be syntactically different but represent the same ground value. Our definition requires that $t_1[B]$ be an ancestor of $t_2[B]$ (or vice-versa), which means they represent the same ground value.  This definition extends to FDs $X\rightarrow Y$ with a set of attributes $X,Y$ in $\mc{R}$. The definition} reduces to the classical FD consistency when $R'$ is ground. Assuming \vghs have fixed size, consistency checking in the presence of generalized values is in quadratic time \blue{according to} $|R'|$.

\blue{\begin{example}\em \label{ex:ext-consistency} Consider an FD $\varphi:[\attr{GEN}, \attr{DIAG}]\rightarrow [\attr{MED}]$ that states two patients of the same gender that are diagnosed with the same disease should be prescribed the same medication. According to the classic definition of FD consistency,  $t_1$ and $t_2$ in Table~\ref{tab:target} are inconsistent as $t_1$ and $t_2$ are males with osteoarthritis, i.e.  $t_1 [\attr{GEN,DIAG}]=t_2 [\attr{GEN,DIAG}]$ $=$ \{\val{male}, \val{osteoarthritis}\}, but are prescribed different medications, i.e. $t_1[\attr{MED}] = $ \val{ibuprofen} $\ne$\val{intropes}  $ = t_2[\attr{MED}]$. Under our revised definition of consistency for a generalized relation, if we update $t_2[\attr{MED}]$ to \val{NSAID}, which is the ancestor of \val{ibuprofen}, then $t_1$ and $t_2$ are now  consistent.  In our revised consistency definition,  \ignore{We have $t_1 [\attr{GEN,DIAG}] = t_2 [\attr{GEN},\attr{DIAG}]$, and  $t_1[\attr{MED}] = $ \val{ibuprofen} $\sqsubseteq$ \val{NSAID} $= t_2[\attr{MED}]$ (cf. Fig~\ref{fig:med} (b) for the \vgh of \attr{MED}).}  $t_1[\attr{MED}] = $ \val{ibuprofen} $\preceq$ \val{NSAID} $= t_2[\attr{MED}]$, indicating that the general value $t_2[\attr{MED}]=$ \val{NSAID} may represent the ground value $t_1[\attr{MED}]=$ \val{ibuprofen}. However, if we were to update  $t_2[\attr{MED}]$ to \val{vasodilators}, which is not the ancestor of $t_1[\attr{MED}] = $ \val{ibuprofen}, then $t_1$ and $t_2$ remain inconsistent under the generalized consistency definition, since the general value $t_2[\attr{MED}]=$ \val{vasodilators} cannot represent $t_1[\attr{MED}] = $ \val{ibuprofen}.
\end{example}}

\section{PACAS Overview} \label{sec:pacas}
\label{sec:overview}
We formally define our problem, \blue{and then review the PACAS framework~\cite{huang2018pacas}, including extensions for  generalized repairs.} 

\subsection{Problem Statement} \label{sec:problem}

Consider a client, \client, and a service provider, \service, with databases $D_\client, D_\service$ containing single relations $R_\client, R_\service$, respectively. Our discussion easily extends to databases with multiples relations. We assume a set of FDs $\Sigma$ defined over $R_\client$ that is falsified. We use FDs as the benchmark for error detection, but our framework is amenable to other error detection methods. The shared generalization hierarchies are generated by the service provider (applying the techniques mentioned in Section~\ref{sec:generalization}). The problem of privacy-preserving data cleaning is twofold, defined separately for \client and \service. We assume a generalization level $l_{\max}$, which indicates the maximum level that values in our repaired database can take from the generalization hierarchy.

\noindent \textbf{Client-side:} For every cell $c \in R_\client$ with value $c.{\sf value}$, let $c.{\sf value}^*$ be the corresponding accurate value in $R_\service$. A cell $c$ is considered dirty if $c.{\sf value}\neq c.{\sf value}^*$. We assume the client can initiate a set of requests $r_1,...,r_n$ to $R_\service$ in which each request $r_i$ is of the form $r_i=(t,A,l)$, that seeks the clean value of database cell $t[A]$ at level $l$ in $R_\service$. We assume $\sum_{i}(\nit{price}(r_i)) \le \budget$ for a fixed cleaning budget \budget. Let $R^*_\client$ be the clean version of $R_\client$ where for each cell,  $c.{\sf value}=c.{\sf value}^*$. The problem is to generate a set of requests $r_1,...,r_n$, where the answers (\blue{possibly containing generalized values}) are used to compute a relation $R'_\client$ such that: (i) $R'_\client \models \Sigma$, (ii) $\delta(R'_\client,R^*_\client)$ is minimal, and (iii) for each $c.{\sf value}$, its level $l \le l_{\max}$.

\blue{In our implementation, we check consistency $R'_\client \models \Sigma$ using the consistency definition in Section~\ref{sec:GDBcons}, and measure the distance $\delta(R'_\client,R^*_\client)$  using the semantic distance function $\delta$ (Defn.~\ref{df:distance}).}

\noindent \textbf{Service-side:} The problem is to compute a pricing function $\nit{price}(r_i)$ that assigns a price to each request $r_i$ such that $R_\service$ preserves \xyl-anonymity.

\begin{figure}
  \centering
  \hspace*{-1mm}\includegraphics[width=4.5in]{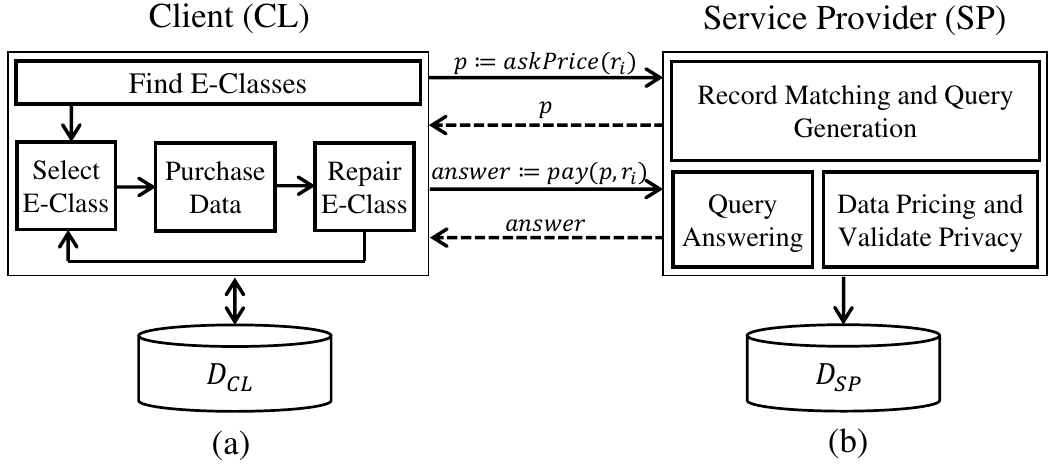}
  \caption{Framework overview.}\label{fig:framework}
\end{figure}

\subsection{Solution Overview}\label{sec:solution}
Figure~\ref{fig:framework} shows the PACAS system architecture consisting of two main units that execute functions for \client and \service. Figure~\ref{fig:framework}(a) shows the \client unit containing four modules. The first module finds equivalence classes in $R_\client$. An {\em equivalence class} (eq) is a set of cells in $R_\client$ with equal values in order for $R_\client$ to satisfy $\Sigma$~\cite{BFFR05}. The next three modules apply an iterative repair for the eqs. These modules select an eq, purchase accurate value(s) of dirty cell(s) in the class, and then repair the cells  using the purchased value.  \blue{If the repairs contain generalized values, the CL unit must verify consistency of the generalized relation against the defined FDs.}  The cleaning iterations continue until $\budget$ is exhausted or all the eqs are repaired.  \blue{We preferentially allocated the budget $\budget$ to cells in an eq based on the proportion of errors in which the cells (in an eq) participate.} Figure~\ref{fig:framework}(b) shows the \service unit.  The \emph{Record Matching} module receives a request $r_i=(t,A,l)$ from \client, identifies a matching tuple $t'$ in $R_\service$, and returns $t'[A]$ at the level $l$ according to the \vgh. The \emph{Data Pricing and Validate Privacy} module computes prices for these requests, and checks whether answering these requests will violate \xyl-anonymity in $R_\service$. If so, the request is not safe to be answered. The \emph{Query Answering} module accepts payment for requests, and returns the corresponding answers to \client. 

\section{Limiting Disclosure of Sensitive Data} \label{sec:disclosure}
The \service must carefully control disclosure of its curated and sensitive data to the \client.  In this section, we describe how the \service services an incoming client request for data, validates whether answering this request is safe (in terms of violating \xyl-anonymity), and the data pricing mechanism that facilitates this interaction.

\subsection{Record Matching and Query Generation} \label{sec:matching}
Given an incoming \client request $r_i=(t,A,l)$, the \service must translate $r_i$ to a query that identifies a matching (clean) tuple $t'$ in $R_\service$, and returns $t'[A]$ at level $l$.  To answer each request, the \service charges the \client a price that is determined by the level $l$ of data disclosure and the adherence of $t'[A]$ to the privacy model.  
To translate a request $r$ to a GQ $G_r$, we assume a schema mapping exists between  $R_\service$ and $R_\client$, with similarity operators $\approx$ to compare the attribute values.  This can be modeled via {\em matching dependencies} (MDs) in  \service~\cite{fan09}. 

\begin{example} \label{ex:query-generation} \em \blue{Consider a request $r\!\!=\!\!(t_2,\attr{MED},l^\attr{MED}_1)$ that requests the medication of the patient in  tuple $t_2$ at level $l^\attr{MED}_1$. To translate $r$ into a generalized query $G_r=\langle Q_r,L_r\rangle$, we use the values of the QI attributes \attr{GEN} and \attr{AGE} in tuple $t_2$.  We define a query $Q_r(R_\service)=\Pi_\attr{MED}(\sigma_{\attr{GEN} = t_2[\attr{GEN}]\wedge \attr{AGE} = t_2[\attr{AGE}]}$ \ $(R_\service))$ requesting the medications of patients in $R_\service$ with the same gender and age as the patient in tuple $t_2$. The GQ $G_r$ level is equal to the level of the request, i.e. $L_r=\{l^\attr{MED}_1\}$. The query $Q_r$ is generated from an assumed matching dependency (MD) $R_\client[\attr{GEN}]= R_\service[\attr{GEN}] \wedge R_\client[\attr{AGE}]= R_\service[\attr{AGE}] \rightarrow R_\client[\attr{MED}]= R_\service[\attr{MED}]$ that states if the gender and age of two records in \service \ and \client \ are equal, then the two records refer to patients with the same prescribed medication.} 



\eatttr{The query matches tuples by comparison between the \attr{GEN},\attr{AGE} attribute values using the similarity operators $\approx_\attr{gen},\approx_\attr{age}$, respectively.} 

\end{example}




\subsection{Enforcing Privacy} \label{sec:pricing}
\blue{Recall from Section~\ref{sec:xyl} that \xyl-anonymity extends \xy-anonymity to consider the attribute domain semantics.} Given a GQ, the \service must determine whether it is safe to answer this query, i.e., decide whether disclosing the value requested in $r_{i}$ violates \xyl-anonymity (defined in Defn.~\ref{df:xyl}).  \blue{In this section, we formalize the privacy guarantees of \xyl-anonymity, and in the next section, review the \palg data pricing algorithm that assigns prices to GQs to guarantee \xyl-anonymity.}

\blue{\xyl-anonymity applies tighter privacy restrictions compared to \xy-anonymity by tuning the parameter $L$ (determined by the data owner).  At higher levels of $L$, it becomes increasingly more difficult to satisfy the \xyl-anonymity condition, while \xyl-anonymity reduces to \xy-anonymity at $l_i = 0$ for every $l_i \in L$. We formally state this in Theorem~\ref{th:privacy}.} \eat{Clearly, \xyl-anonymity reduces to \xy-anonymity when $L$ contains ground levels.}  \xyl-anonymity is a semantic extension of \xy-anonymity.  Similar extensions can be defined for PPDP models such as \xy-privacy, $l$-diversity, $t$-closeness. Unfortunately, all these models do not consider the data semantics in the attribute \dghs and \vghs~\cite{fung}.  

\blue{\begin{theorem}\label{th:privacy} Consider a relation $R$ and let $X$ and $Y$ be two subsets of attributes in $R$. Let $L_1$ and $L_2$ be levels of the attributes in $Y$ with $L_1\le L_2$, i.e. every level of $L_1$ is lower than or equal to its corresponding level in $L_2$. The following holds for relation $R$ with a fixed $k > 1$:

\begin{enumerate}
    \item If $R$ is $(X,Y,L_2)$-anonymous it is also $(X,Y,L_1)$-anonymous, but not vice versa (i.e. $R$ can be $(X,Y,L_1)$-anonymous but not $(X,Y,L_2)$-anonymous).
    \item For any levels $L$ of attributes in $Y$, if $R$ is \xyl-anonymous, it is also \xy-anonymous.
\end{enumerate}

\end{theorem}

\noindent {\bf Proof of Theorem~\ref{th:privacy}.} In the first item, if $R$ is $(X,Y,L_2)$-anonymous, then for every $t$, $|G_1^t(R)| \ge k$ in which $G_1^t=\langle Q^t, L_1\rangle$. Considering $G_2^t=\langle Q^t, L_2\rangle$, we can claim $|G_2^t(R)| \ge k$, which proves $R$ is also $(X,Y,L_1)$-anonymous. The claim holds because if $Q^t$ has $n$ answers in levels of $L_1$, it will have fewer or equal number of answers in the higher level $L_2$ since different values in levels of $L_1$ may be replaced with the same ancestors in the levels of $L_2$. The second item holds because if $R$ is \xyl-anonymous it is also $(X,Y,L^\bot)$-anonymous where $L^\bot$ contains the bottom levels of attributes in $Y$; due to $L^\bot \le L$ and item 1. If $R$ is $(X,Y,L^\bot)$-anonymous, then every value of $X$ appears with $k$ ground values and we can conclude it is \xy-anonymous.\boxtheorem}\vspace{2mm}


\eat{ 
The \xyl-anonymity model allows users to apply tighter privacy restrictions compared to \xy-anonymity by tuning the value of $l$. A higher level $l$ makes it harder to satisfy the \xyl-anonymity condition while \xyl-anonymity reduces to \xy-anonymity if $l$ is the bottom level.
}


\subsection{Pricing Generalized Queries.} \label{sec:gprice} 
PPDP models have traditionally been used in non-interactive settings where a privacy-preserving table is published once.  We take a user-centric approach and let users dictate the data they would like published.  We apply PPDP in an interactive setting where values from relation $R_{\service}$ are published incrementally according to (user) \client requests, while verifying that the disclosed data satisfies \xyl-anonymity~\cite{huang2018pacas}. We adopt a data pricing scheme that assigns prices to GQs by extending the baseline data pricing algorithm defined in~\cite{deep} to guarantee \xyl-anonymity.  We review this baseline pricing algorithm next.

\noindent \textbf{Baseline Data Pricing.} \hspace{0.2cm}
The baseline pricing model computes a price for a query $Q$ over relation $R$ according to the amount of information revealed about $R$ when answering $Q$~\cite{deep}.

\begin{figure}
  \centering
  \includegraphics[width=2in]{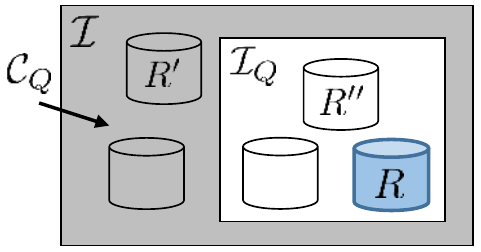}
  \caption{Possible relations $\mc{I}$, conflict set  $\mc{C}_Q$ and admissible relations $\mc{I}_Q$ for query $Q$. 
  }
  \label{fig:conflict}
\end{figure}

Given a query $Q$ over a relation $R$ (a database with single relation $R$), the baseline pricing model determines the price of $Q$ based on the amount of information revealed about $R$ by answering $Q$. Let $\mc{I}$ be a set of {\em possible relations} that the buyer believes to be $R$, representing his initial knowledge of $R$.  As the buyer receives answers to $Q(R)$, he gains new knowledge, allowing him  to eliminate relations $R'$ from $\mc{I}$, which  provide a different answer $Q(R') \neq Q(R)$. This set of eliminated instances $R'$ is called the {\em conflict set of $Q$}  denoted as $\mc{C}_Q$, and intuitively represents the amount of information that is revealed by answering $Q$ (Figure~\ref{fig:conflict}).  As more queries are answered, the size of $\mc{I}$ is reduced.  We can apply a set function that uses $\mc{C}_Q$ to compute a price for $Q$.  We can use the {\em weighted cover set} function with predefined weights assigned to the relations in $\mc{I}$.  Query prices are computed by summing the weights for relations in $\mc{C}_Q$, which has been shown to give arbitrage-free prices~\cite{deep}.  In practice, the set $\mc{I}$ is usually infinite making it infeasible to implement.  To circumvent this problem, a smaller, finite subset $\mc{S}$ called the {\em support set} is used to generate arbitrage-free prices~\cite{deep}. The support set is defined as the neighbors of $R$, generated from $R$ via tuple updates, insertions, and deletions. The values that are used to generate the support set are from the same domain of the original relation $R$.

\begin{algorithm}[h]
\SetKwInOut{Input}{Input}
\SetKwInOut{Output}{Output}
\AlgoDisplayBlockMarkers\SetAlgoBlockMarkers{}{end}
\Input{A query $Q$, a database $D$, support set $\mc{S}$, weight function $w$} 
\Output{Price to answer $Q$}

$p\leftarrow  0$;

\For{$D' \in \mc{S}$}{
    \lIf{$Q(D')\neq Q(D)$}{
        $p\leftarrow p+w(D')$\label{ln:d1}
    }
}
\KwRet{$p$};
\caption{$\nit{price} (Q,D,\mc{S},w)$~\cite{deep}}
\label{alg:pricing}
\end{algorithm}

Several optimizations are applied to the baseline pricing, and its effectiveness is experimentally justified~\cite{deep}. Most importantly, the relations in the support set can be modeled by update operations. That is, we generate each relation in $\mc{S}$ from $R$ by applying its corresponding update operation and use the resulting relation to compute the value of the weighted function as the final price. We roll-back the update to restore $R$, and continue this process to compute the weighted function \blue{using} other relations in $\mc{S}$.  We avoid storing all databases in $\mc{S}$ to enable more efficient price computations.



Algorithm~\ref{alg:pricing} provides pseudocode of the baseline algorithm.  The algorithm takes query $Q$, database $D$,  a support set $\mc{S}$, and a weight function $w$, and computes the price to answer $Q$ over $R$. The baseline algorithm is history-aware as input $\mc{S}$ excludes databases that were already considered by past queries. 


\noindent \textbf{\palg Algorithm.} \hspace{0.2cm}
\blue{We describe the \palg algorithm (originally introduced in~\cite{huang2018pacas}) that enforces \xyl-anonymity over  $R$ (equivalently $R_\service$ in our framework).}  We first present the definition of a \emph{safe query}, i.e., criteria for a GQ to preserve \xyl-anonymity. 

\begin{definition}  \label{df:protect} \em (Safe Query) Consider a GQ $G$ over a relation $R$ with schema $\sch{R}$, $X,Y \subseteq \sch{R}$, and levels $L$ corresponding to attributes in $Y$. Let $\mc{I}_G \subseteq \mc{I}$ be the set of relations $R''$ such that $G(R)=G(R'')$. $G$ is safe (or preserves \xyl-anonymity of $R$) with value $k$, if for every tuple $t \in R$, there are at least $k$ tuples in the set of answers $\{t''\;|\; \exists R'' \in \mc{I}_G, t'' \in \Qry{G}{t}(R'')\}$ where $\Qry{G}{t}=\langle \Qry{Q}{t},L\rangle$ is a GQ with $\Qry{Q}{t}(R'')=\Pi_{Y}(\sigma_{X=t[X]}(R''))$.\end{definition}


In Definition~\ref{df:protect}, $\mc{I}_G$ represents the set of relations that the buyer believes $R$ is drawn from after observing the answer $G(R)$. If there are at least $k$ tuples in the answer set of $\Qry{G}{t}$ over $\mc{I}_G$, this indicates that the buyer does not have enough information to associate the values in $X$ to less than $k$ values of $Y$ at  level $L$, thus preserving \xyl-anonymity.  If the \service determines that a GQ is safe, he will assign a finite price relative to the amount of disclosed information about $R$. 


Algorithm~\ref{alg:ppricing} presents the \palg details. The given support set $\mc{S}$ represents the user's knowledge about relation $R$ after receiving the answer to past purchased queries. We maintain a set $\mc{S}_G$ that represents all admissible relations after answering $G$ and captures the user's posterior knowledge about $R$ after answering $G$. We use $\mc{S}_G$ to check whether answering $G$ is safe. This is done by checking over instances in $\mc{S}_G$ whether values in $X$ are associated with at least $k$ values of $Y$ using the query $\Qry{G}{t}$. The price for a query is computed by summing the weights of the inadmissible relations in the conflict set $\mc{C}_G = \mc{S} \setminus \mc{S}_G$ (Line~\ref{ln:price}).  Similar to the baseline pricing, we use the support set $\mc{S}$ and admissible relations $\mc{S}_G$ rather than $\mc{I}$ and $\mc{I}_G$, respectively. We iterate over tuples $t \in R$ (Line~\ref{ln:for}), and check whether values in $X$ are associated with less than $k$ values in $Y$ over relations in $\mc{S}_G$.   If so, we return an infinite price reflecting that query $G$ is not safe (Line~\ref{l:check}).



\begin{proposition} \em
If the input GQ is not safe, Algorithm~\ref{alg:ppricing} preserves \xyl-anonymity by returning an infinite price. \end{proposition}

\noindent \emph{Proof sketch:} According to Definition~\ref{df:protect}, if $G$ violates \xyl-anonymity then for $t \in R$, there are less than $k$ answers to $\Qry{G}{t}$ over relations in $\mc{I}_G$. Since $\mc{S}_G \subseteq \mc{I}_G$, there will be less than $k$ answers to $\Qry{G}{t}$ over relations in $\mc{S}_G$, meaning Algorithm~\ref{alg:ppricing} assigns infinite price to $G$ in Line~\ref{l:check}.  Note that if Algorithm~\ref{alg:ppricing} returns an infinite price, it does not imply $G$ is unsafe (this only occurs when $\mc{S}=\mc{I}$). \qedsymbol



\begin{algorithm}[h]
\SetKwInOut{Input}{Input}
\SetKwInOut{Output}{Output}
\Input{$G$, $R$, $\mc{S}$, $w$}
\Output{Price of $G$\ignore{ if it is safe w.r.t. \xyl-anonymity}}

$p\leftarrow 0$; $\mc{S}_G\leftarrow 0$;

\For{$T \in \mc{S}$}{
    \lIf{$G(T)=G(R)$}{                
        $\mc{S}_G \leftarrow \mc{S}_G \cup \{T\}$
    } \lElse {
    	$p\leftarrow p+w(T)$\label{ln:price}
    }
}

\For{$t \in R$ \label{ln:for}}{
	$A \leftarrow  \emptyset$;
    
	\lFor{$R'' \in \mc{S}_G$}{        
        $A \leftarrow  A \cup \Qry{G}{t}(R'')$\label{ln:d1}
    }
    
    \lIf{$|A| < k$}{\KwRet{$\infty$}\label{l:check}}
}

\KwRet{$p$};
\caption{$\palg(G,R,\mc{S},w)$}
\label{alg:ppricing}
\vspace{-1mm}
\end{algorithm}

Given the interactive setting between the \client and the \service, we must ensure that all (consecutively) disclosed values guarantee \xyl-anonymity over $R$.   We ensure that \palg is history-aware by updating the support set $\mc{S}$ after answering each GQ.  The \service uses the pricing function in Algorithm~\ref{alg:ppricing} to update $\mc{S}$ to reflect the current relations $R'$ that have been eliminated by answering the latest GQ.  The \service \ implements $\nit{AskPrice}(r_i, R_\service)$ (cf. Figure~\ref{fig:framework}) by  translating $r_i$ to a GQ, and then invoking \palg.

We assume that requesting a query price is free. We acknowledge that returning prices might leak information about the data being priced and purchased. This problem is discussed in the data pricing literature, particularly to incentivize data owners to return trustful prices when prices reveal information about the data (cf.~\cite{roth} for a survey on this issue). We consider this problem as a direction of future work.


\subsection{Query Answering}

A \client data request is executed via the $\nit{Pay}(p,r_i,R_\service)$ method, where she purchases the value in $r_i$ at price $p$. The \emph{Query Answering} module executes $\nit{Pay}(p,r_i,R_\service)$ via \service accepts payment,  translates $r_i$ to a GQ, and returns the answer over $R_\service$ to \client. Lastly, \service updates the support set $\mc{S}$ to ensure \palg has an accurate history of disclosed data values.

We note that all communication between \client and \service is done via the  \nit{AskPrice} and \nit{Pay} methods (provided by \service).   We assume there is a secure communication protocol between the \client and \service, and that all data transfer is protected and encrypted. The model is limited to a single \service that sells data at non-negotiable prices.  We intend to explore general cases involving multiple \service providers and price negotiation as future work.

\section{Data Cleaning with Generalized Values} 
\label{sec:framework}

Existing constraint-based data repair algorithms that propose updates to the data to satisfy a set of data dependencies assume an open-access data model with no data privacy restrictions  \cite{BFFR05,CM11a,BIG10,BIGG13,NADEEF}.  In these repair models, inadvertent data disclosure can occur as the space of repair candidates is not filtered nor transformed to obscure sensitive values.  A privacy-aware data repair algorithm must address the challenge of providing an accurate repair  to an error while respecting data generalizations and perturbations to conceal sensitive values.

\blue{In earlier work, we introduced \calg, a data repair algorithm that resolves errors in a relation $R_\client$ using data purchased from a service provider $R_\service$~\cite{huang2018pacas}.}  The key distinctions of \calg from past work include: (i) \calg interacts with the service provider, \service, to purchase trusted values under a constrained budget $\budget$.  This eliminates the overhead of traversing a large search space of repair candidates; and (ii) \calg tries to obtain values with highest utility from \service for repairing \client.  \blue{However, the existing \calg algorithm grounds all purchased values thereby not allowing generalized values in $R_\client$.  In this paper, we extend \calg to consider generalized values in $R_\client$ by re-defining the notion of consistency between a relation and a set of FDs (Section~\ref{sec:GDBcons}), and revising the budget allocation algorithm to requests according to the number of errors in which cells in an equivalence participate (Section~\ref{sec:rgen}).  In contrast, the existing \calg provides fixed budget allocations.}  We give an overview of our cleaning algorithm and subsequently describe each component in detail. 

\subsection{Overview}
For a fixed number of FDs $\Sigma$, the problem of finding minimal-cost data repairs to $R_\client$  such that $R_\client$ satisfies $\Sigma$ is  NP-complete \cite{BFFR05}.  Due to these intractability results, we necessarily take a greedy approach that cleans cell values in $R_\client$ that maximally reduce the overall number of errors w.r.t. $\Sigma$. This is in similar spirit to existing techniques that have used various weighted cost functions \cite{BFFR05,chiang14} or conflict hypergraphs \cite{BIGG13} to model error interactions among data dependencies in $\Sigma$. 

Given $R_\client$ and $\Sigma$, we identify a set of error cells that belong to tuples that falsify some $\sigma \in \Sigma$.  For an error cell $e \in \mathcal{E}$, we define eq($e$) as the equivalence class to which $e$ belongs.  An equivalence class is a set of database cells with the same value such that $\Sigma$ is satisfied~\cite{BFFR05}.   We \blue{use} eqs for two reasons: (i) by clustering cells into eqs, we determine a repair value for a \emph{group of cells} rather than an individual cell, thereby improving performance; and (ii) we utilize every cell value  within an eq to find the best repair.

\eat{In Algorithm~\ref{alg:driver}, we clean an error cell w.r.t. a group of cells that must have the same value according to some FDs. The set of these cells is called an \emph{equivalence class}~\cite{BFFR05}, {\em eq} in short, that we will review below. } 
The \calg algorithm repairs FD errors by first finding all the eqs in $R_\client$. The algorithm then iteratively selects an eq, and purchases the true value of a cell, and updates all dirty cells in the same class to the purchased value.  \blue{The \service may decide that returning a generalized value is preferred to protect user sensitive data in $R_\service$.  If so, then the \client must validate consistency of its data including these generalized values.} At each iteration, we repair the eq class with cells that participate in the largest number of FD errors. At each iteration, \calg assigns a portion of the budget \budget that is proportional to the number of errors relative to the total number of errors in $R_\client$ \blue{(this new extension to \calg is discussed in Section~\ref{sec:rgen})}. \calg continues until all the eqs are repaired, or the budget is exhausted.


\begin{algorithm}[h]
\SetKwInOut{Input}{Input}
\SetKwInOut{Output}{Output}
\SetKw{Break}{break}
\Input{$R_\client$, $R_\service$, $\Sigma$, $\budget$}
\Output{Clean $R_\client'$}
$R_\client'\leftarrow R_\client$;\\
$\nit{EQ} \leftarrow\nit{GenerateEQs}(R_\client',\Sigma)$;\label{l:eclass}\\
\blue{$B \leftarrow \budget$;}\\
\For{$\nit{eq} \in \nit{EQ}$}{\lIf{$\nit{Resolved}(\nit{eq})$}{$\nit{EQ} \leftarrow \nit{EQ} \setminus \{\nit{eq}\}$}\label{l:filter}}
\While{$B > 0$ {\bf and} $\nit{EQ} \neq \emptyset$}{
    $\nit{eq} \leftarrow \nit{Select}(\nit{EQ});$\label{l:mosterror}\\
    $r_i \leftarrow \nit{GenerateRequest}(\nit{eq}, \alpha_i \times B, R_\service);$\label{l:bestr}\\
    \If{$r_i \neq \Null$}{
    $p_i \leftarrow \nit{AskPrice}(r_i, R_\service);$\label{l:askp}\\
    $u_i  \leftarrow \nit{Pay}(p_i, r_i, R_\service)$;\\
    $B \leftarrow B - p_i$;\\
	$\nit{ApplyRepair}(\nit{eq},u_i)$;\label{l:repair}} 
	$\nit{EQ} \leftarrow \nit{EQ} \setminus \nit{eq}$;\label{l:removec}
}
\KwRet{$R_\client'$}
\caption{$\nit{SafeClean}(R_\client,R_\service,\Sigma,\budget)$}
\label{alg:driver}
\end{algorithm}

\noindent \textbf{SafeClean Algorithm.} Algorithm~\ref{alg:driver} gives details of SafeClean's overall execution.  
The algorithm first generates the set of eqs via \nit{GenerateEQs} in Line~\ref{l:eclass}. Equivalence classes containing only one value are removed since there is no need for repair. \eat{$\nit{EQ}$ that have only one value in its cells and therefore does not need repair. In each cleaning iteration, the algorithm selects an equivalence class $\nit{eq}_i$ by \nit{Select} and repairs its cells (Line~\ref{l:mosterror}).}   In Line~\ref{l:mosterror}, \calg selects an equivalence class $\nit{eq}_i$ with cells participating in the largest number of violations for repair (further details in Section~\ref{sec:err}). To repair the error cells in $\nit{eq}_i$, \calg generates a request $r_i$ using \nit{GenerateRequest} that requests a repair value for a cell in $\nit{eq}_i$ (Line~\ref{l:bestr}). This request is made at the lowest possible level (less than $l_{max}$) at a price allowable within the given budget. The algorithm assigns a fraction of the remaining budget, i.e. $\alpha_i \times B$ ($B \le \budget$ is the remaining budget) to purchase data at each iteration. This fraction depends on the number of violations in $\nit{eq}_i$ (cf. Section~\ref{sec:rgen} for  details). If such a request can be satisfied, the value(s) are purchased and applied (Lines~\ref{l:askp}-\ref{l:repair}).  If there is insufficient budget remaining to purchase a repair value, then \eat{Otherwise and if the budget is not enough for purchasing any value for the cells in $\nit{eq}_i$ ($r_i$ is \Null),} the eq cannot be repaired.  In either case, \calg  removes $\nit{eq}_i$ from $\nit{EQ}$ and continues with the next eq (Line~\ref{l:removec}). \calg terminates when $\budget$ is exhausted, or there are no eqs are remaining.   We present details of eq generation, selection, request generation, and data purchase/repair in the following sections.

\subsection{Generating Equivalence Classes} 

The eqs are generated by \nit{GenerateEQs} in Algorithm~\ref{alg:eq} that takes as input $R_\client$ and $\Sigma$, and returns the set of eqs $\nit{EQ}$. For every cell $c_i \in R_\client$, the procedure initializes the eq of $c_i$ as $\nit{eq}(c_i)=\{c_i\}$, and adds it to the set of eqs $\nit{EQ}$ (Line~\ref{l:init}). We then iteratively merge the eqs of any pair of cells $c_1=t_1[B],c_2=t_2[B]$ if there is a FD $\varphi: A \rightarrow B \in \Sigma$, $t_1[A]=t_2[A]$, and both $t_1[A]$ and $t_2[A]$ are ground. The procedure stops and returns $\nit{EQ}$ when no further pair of eqs can be merged.

\begin{algorithm}[h]
\SetKwInOut{Input}{Input}
\SetKwInOut{Output}{Output}
\SetKw{Break}{break}
\AlgoDisplayBlockMarkers\SetAlgoBlockMarkers{}{end}
\Input{$R_\client$, $\Sigma$,}
\Output{The set of equivalence classes $\nit{EQ}$}

$\nit{EQ} \leftarrow \emptyset$;

\lFor{$c_i \in R_\client$}{
    $\nit{EQ} \leftarrow \nit{EQ} \cup \{ \{c_i\} \}$\label{l:init}
}

\For{{\bf every} $t_1,t_2 \in R_\client$ {\bf and} $\varphi:A\ra B \in \Sigma$}{
    \lIf{$t_1[A]=t_2[A]$ }{
        $\nit{merge}(\nit{EQ}(t_1[B]), \nit{EQ}(t_2[B]))$
    }
}

\KwRet{$\nit{EQ}$;}

\caption{$\nit{GenerateEQs}(R_\client,\Sigma)$}
\label{alg:eq}
\end{algorithm}

\begin{example}\em
Given FD $\varphi:[\attr{GEN}, \attr{DIAG}]\rightarrow [\attr{MED}]$ in Table~\ref{tab:target}, since $t_1$, $t_2$ and $t_3$ have the same attribute values on \attr{GEN} and \attr{DIAG}, we merge $t_1[\attr{MED}]$, $t_2[\attr{MED}]$ and $t_3[\attr{MED}]$ into the same $\nit{EQ}_1$.   Similarly, we cluster $t_4$ and $t_5$ into the same $\nit{EQ}_2$.   
\label{ex:eclass}
\end{example}


\subsection{Selecting Equivalence Classes for Repair} \label{sec:err}
In each iteration of Algorithm~\ref{alg:driver}, we repair the cells in an eq $\nit{eq}_i$ that will resolve the most errors in $R_\client$ w.r.t. $\Sigma$. \eat{Here, an {\em error} w.r.t. an FD $\varphi:(A \ra B) \in \Sigma$ is a pair of tuples $\{\tp{t_1},\tp{t_2}\} \subseteq R_\client$ that falsifies $\varphi$ according to FD semantics in Section~\ref{sec:GDBcons}.} To achieve this goal, we choose $\nit{eq}_i$ as the eq with cells participating in the largest number of errors. For a cell $c_j \in R_\client$ and an FD $\varphi: A \rightarrow B \in \Sigma$, let $\errors(R_\client, \varphi, c_j)$ be the set of errors $\{\tp{t_1},\tp{t_2}\}$ w.r.t. $\varphi$ and $c_j \in \{t_1[A], t_1[B], t_2[A], t_2[B]\}$. For an eq $\nit{eq}_i$, let $\errors(R_\client, \varphi, \nit{eq}_i)=\bigcup_{c_j \in \nit{eq}_i} \errors(R_\client, \varphi, c_j)$ and $\errors(R_\client, \Sigma, \nit{eq}_i)=\bigcup_{\varphi \in \Sigma} \errors(R_\client, \varphi, \nit{eq}_i)$. The eq $\nit{eq}_i$ returned by \nit{Select} in Line~\ref{l:eclass} of Algorithm~\ref{alg:driver} is the eq with the largest number of errors in $\errors(R_\client, \Sigma, \nit{eq}_i)$. In other words, this is the number of tuple pairs that every cell in $\nit{eq}_i$ participates, summed over all FDs in $\Sigma$.

\begin{example}\label{ex:ccheck} \em We continue Example~\ref{ex:eclass} with $\nit{EQ}_1$ and $\nit{EQ}_2$. According to our error definition, given FD $\varphi:[\attr{GEN}, \attr{DIAG}]\rightarrow [\attr{MED}]$, $\nit{EQ}_1$ is involved in three errors with tuples  ($\{t_1,t_2\}$, $\{t_1,t_3\}$ and $\{t_2,t_3\}$), while    $\nit{EQ}_2$ has one error ($\{t_3,t_4\}$).  Hence, our algorithm will select $\nit{EQ}_1$ to repair first.   
\end{example}

\subsection{Data Request Generation} \label{sec:rgen}
\eat{The budget allocation algorithm in \cite{huang2018pacas} only considered the number of errors in the relation. In this paper, we also consider the number of FD violations each dirty cell involved and the total number of erros in the relation. We propose a new budget allocation algorithm to make full use of the user-given budget for dirty cells. }

To repair cells in $\nit{eq}_i$, we generate a request $r_i=\langle c, l\rangle$ by \nit{GenerateRequest} (Algorithm~\ref{alg:best}) that requests accurate and trusted value of a cell $c \in \nit{eq}_i$ at level $l$ from $R_\service$.  There are two restrictions on this request: (i) the price must be within the budget $\alpha_i \times \budget$, and (ii) the level $l$ must be $\le l_\nit{max}$. The value $\alpha_i$ is defined as follows: 

\vspace{-1mm}
$$\alpha_i=\frac{\errors(R_\client, \Sigma, \nit{eq}_i)}{\sum_{\nit{eq}_j \in \nit{EQ}}{\errors(R_\client, \Sigma, \nit{eq}_j)}}.$$\vspace{-1mm}

\noindent In this definition, the allocated budget $\alpha_i \times B$ to each iteration is proportional to the number of FD violations in $\nit{eq}_i$, and also depends on the total number of errors in $R_\client$. This allocation model improves upon previous work that decreases the budget allocated to the $i^{th}$-request by a factor of $\frac{1}{i}$, and does not adjust the allocation to the number of errors in which a cell participates~\cite{huang2018pacas}. Note that if the price paid for $r_i$ (i.e. $p_i$) is less than this allocated budget, the remaining budget carries to the next iteration through $B$.



If there is no such request, \nit{GenerateRequest} returns null, indicating that $\nit{eq}_i$ cannot be repaired with the allocated budget (Line~\ref{l:null}). If there are several requests that satisfy (i) and (ii), we follow a greedy approach and select the request at the lowest level with the price to break ties (Line~\ref{l:tie}). For a cell $c_i \in \nit{eq}_i$, \nit{LowestAffordableLevel} in Line~\ref{l:afford} finds the lowest level in which the value of $c_i$ can be purchased from $R_\service$ considering the restrictions in (i) and (ii).  Our greedy algorithm spends most of the allocated budget $\alpha_i \times B$ in the current iteration. An alternative approach is to select requests with the highest acceptable level ($l_\nit{max}$) to preserve the budget for future iterations.

\begin{algorithm}[h]
\SetKwInOut{Input}{Input}
\SetKwInOut{Output}{Output}
\SetKw{Break}{break}
\AlgoDisplayBlockMarkers\SetAlgoBlockMarkers{}{end}
\Input{eq $C$, budget $b$ and $R_\service$.}
\Output{Request $r_i$}

$l \leftarrow l_h$;
$p \leftarrow \infty$;
$c \leftarrow \Null$;

\For{$c_i \in C$}{
    $l_i \leftarrow \nit{LowestAffordableLevel}(c_i,b,l_\nit{max},R_\service)$;\\ \label{l:afford}
    $p_i \leftarrow \nit{AskPrice}(\langle c_i, l_i\rangle, R_\service)$;\\
    \If{$l_i < l$ {\bf or} $(l_i == l$ {\bf and} $p_i \le p)$\label{l:tie}}{$c \leftarrow c_i$;$\;\;l \leftarrow l_i$;$\;\;p \leftarrow p_i$;}
}
\lIf{$c \neq \Null$}{\KwRet{$\langle c, l\rangle$}}{\KwRet{\Null};\label{l:null}}

\caption{$\nit{GenerateRequest}(C,b,R_\service)$}
\label{alg:best}
\end{algorithm}

\subsection{Purchase Data and Repair}\label{sec:gquery}

To repair the cells in $\nit{eq}_i$, Algorithm~\ref{alg:driver} invokes $\nit{Pay}(r_i,R_\service)$ to purchase the trusted value $u_i$ and replaces the value of every cell in $\nit{eq}_i$ with $u_i$ in \nit{ApplyRepair}. The algorithm then removes the eq $\nit{eq}_i$ from $\nit{EQ}$ and continues to repair the next eq.  The algorithm stops when there is no eq to repair or the budget is exhausted.
\begin{example}\em
Continuing from Example~\ref{ex:ccheck}, since $\nit{EQ}_1$ has the largest number of errors, we purchase the trusted value from $R_\service$ to repair  $\nit{EQ}_1$ first. If the purchased value is general value \val{NSAID}, we update all cells in $\nit{EQ}_1$, i.e., $t_1[\attr{MED}]$, $t_2[\attr{MED}]$ and $t_3[\attr{MED}]$ to \val{NSAID} to resolve the inconsistency. 
\end{example}

\vspace{-3mm}
\subsection{Complexity Analysis} \label{sec:complexity}
We analyze the complexity of \calg's modules:


\begin{itemize}[nolistsep, leftmargin=*]
\item \blue{Identifying errors involves computing equivalence classes, and selecting an equivalence class for repair.}  \eat{Error identification is based on selecting the dirtiest cell $e$, and resolved w.r.t. an equivalence class.}  In the worst case, this is quadratic in the number of tuples in $R_\client$. 
\item For each data request to resolve an error, executing \palg and $pay$ rely on $R_\service$, the support set $\mc{S}$, and the complexity of GQ answering. We assume the size of $\mc{S}$ is linear w.r.t the size of $R_\service$. The complexity of running GQs is the same as running SQL queries. Thus, all procedures run in polynomial time \blue{to} the size of $R_\service$.
\item \blue{In \emph{applyRepair}, for each returned value from \service, we \eat{must ground}update the error cells in $R_\client$ for each equivalence class \blue{and} each FD, taking time on the order of $\mathcal{O}(|\mathcal{E}||A||\Sigma|)$  for attribute domain size $|A|$.}  We must update the affected cells in the equivalence class, and their dirty scores; both taking time bounded by the number of cells in $R_\client$. Hence, Algorithm~\ref{alg:driver} runs in polynomial time \blue{to} the size of $R_\service$ and $R_\client$.
  
\end{itemize}

\eatttr{
 $N^2 \cdot |A| \cdot |\sigma| \cdot pricing$
}

\section{Experiments} \label{sec:experiments}

Our evaluation focuses on the following objectives: 
\begin{enumerate}[nolistsep,leftmargin=*]

\eat{data repair accuracy. Recall that for the given query, the price of data is determined by the support set, and increasing the size of support size will allow more specific query to be answered at lower level, which will also increase the running time.}  

\item \blue{We evaluate the impact of generalized values on the repair error, and the runtime performance.  In addition, we study the proportion of generalized repair values that are recommended for varying budget levels.}
\item The efficiency and effectiveness of \palg to generate reasonable prices that allow \calg to effectively repair the data.

\item We evaluate the \blue{repair error} and scalability of \calg as we vary the parameters $k, l, e$ and $\mc{B}$ \blue{to study the repair error to runtime tradeoff.}

\item \blue{We compare \calg against PrivateClean, a framework that explores the link between differential privacy and data cleaning~\cite{krishnan}.  We study the repair error to runtime tradeoff between these two techniques, and show that data randomization in PrivateClean significantly hinders error detection and data repair, leading to increased repair errors.} 


\end{enumerate}

\subsection{Experimental Setup}
\label{sec:expsetup}
\blue{We implement \pacas \ using Python 3.6 on a server with 32 Core Intel Xeon 2.2 GHz processor with 64GB RAM.  We describe the datasets, baseline comparative algorithm, metrics and parameters.  The datasets, their schema, and source code implementation can be found at~\cite{code}.  
}

\eat{over attributes "drug name", "condition" on Clinical Trials Data set; "total person income", "education" on Census dataset; "address", "violation description"(the food sanitation violation such as unapproved-food, unknown-source, etc) on Food Inspection dataset.}

\eat{To generate these \vgh of the given dataset, we first extract all of the domain value of the datasets, and manually check them with authenticated external ontology sources such as the Bioprotal Medical Ontology \cite{medical_ontology}, University of Maryland Disease Ontology \cite{disease_ontology}, Libraries of ontologies from University of Michigan \cite{ontology}, Personal income class of U.S. \cite{income_ontology}, Structure of U.S. Education \cite{education_ontology},  etc and Food Service Establishment Inspection Code \cite{food_ontology}. With these authenticated external ontology sources, we can identify the level, ancestor and descendants of each value in our datasets to build the \vgh. The degree of \vgh complexity is sorted as Clinical $>$ Census $>$ Food because of the intrinsic characteristics of the datasets and their domain values. Specifically, the average number of attributes per level on the \vgh of Clinical around 8 while it is 7 and 4 on Census and Food respectively.}


\begin{table}[!tb]
\small{
    \begin{minipage}{.5\linewidth}
            \caption{\blue{Data characteristics.}}
\begin{tabular}{l|cccc}
\toprule
{\bf }  & {\bf Clinical}		& {\bf Census} & {\bf Food} 	\\
\midrule
$|R_\client|$		&	345,000		&		300,000	& 	30,000	 \\
$n$		        	&		29		&		40	& 	    11   	\\
$|\Pi_{A}(R)|$		&	61/73	&		17/250   & 	248/70	 \\
$|\vgh^A|$	    	& 	5/5	&	 	5/6	&   5/5	 \\
\bottomrule
\end{tabular}

        \label{tb:dataset}
    \end{minipage}%
    \begin{minipage}{.5\linewidth}
            \caption{\blue{Parameter values (defaults in bold).}}
\begin{tabular}{ | l | l | l |}
          \hline
          \textbf{Sym.} & \textbf{Description} & \textbf{Values} \\
          \hline
        $\mc{B}$  & budget  & 0.2, 0.4, 0.6, \textbf{0.8} \\
        \hline
          $|\mathcal{S}|$  & support set size  & 6, 8, \textbf{10}, 12, 14 \\
          \hline
          $l$ & generalization level & 0, 1, \textbf{2}, 3, 4\\
          \hline
           $k$ & \#tuples in X-group & 1, 2, \textbf{3}, 4, 5 \\ 
          \hline
          $e$  & error rate & 0.05, \textbf{0.1}, 0.15, 0.2, 0.25 \\
          \hline
\end{tabular}

        \label{tbl:defaults}
    \end{minipage} 
    }
\end{table}

\ignore{
\begin{table}
\small{
\centering
\caption{\blue{Data characteristics.} \fei{Yu: fill-in, and put this table in a row with param. table, should fit.}}
\label{tb:dataset}
\vspace{-4mm}
\begin{tabular}{l|cccc}
\toprule
{\bf }  & {\bf Clinical}		& {\bf Census} & {\bf Food} 	\\
\midrule
$|R_\client|$		&	345,000		&		300,000	& 	30,000	 \\
$n$		        	&		29		&		40	& 	    11   	\\
$|\Pi_{A}(R)|$		&	\tbf/\tbf	&		\tbf/\tbf   & 	\tbf/\tbf	 \\
$|\vgh^A|$	    	& 	\tbf/\tbf	&	 	\tbf/\tbf	&   \tbf/\tbf	 \\
\bottomrule
\end{tabular}
}
\end{table}

\begin{table}
\small{
\caption{\blue{Parameter values (defaults in bold).} \label{tbl:defaults}}
\vspace{-4mm}
\begin{tabular}{ | l | l | l |}
          \hline
          \textbf{Sym.} & \textbf{Description} & \textbf{Values} \\
          \hline
        $\mc{B}$  & budget  & 0.2, 0.4, 0.6, \textbf{0.8} \\
        \hline
          $|\mathcal{S}|$  & support set size  & 6, 8, \textbf{10}, 12, 14 \\
          \hline
          $l$ & generalization level & 0, 1, \textbf{2}, 3, 4\\
          \hline
           $k$ & \#tuples in X-group & 1, 2, \textbf{3}, 4, 5 \\ 
          \hline
          $e$  & error rate & 0.05, \textbf{0.1}, 0.15, 0.2, 0.25 \\
          \hline
\end{tabular}
}
\end{table}
}
\noindent \textbf{Datasets.}
\blue{We use three real datasets. Table~\ref{tb:dataset} gives the data characteristics, showing a range of data sizes in the number of tuples ($|R_\client|$), number of attributes ($n$), number of unique values in the sensitive attribute $A$ ($|\Pi_{A}(R)|$), and the  height of the attribute $\vgh^A$ ($|\vgh^A|$).} We denote sensitive attributes with an asterisk (*).

\noindent \uline{Clinical Trials (Clinical).}   The Linked Clinical Trials database describes patient demographics, diagnosis, prescribed drugs, symptoms, and treatment~\cite{clinic}.  We select the \attr{country}, \attr{gender}, \attr{source} and \attr{age} as QI attributes.  We define two FDs: (i) $\varphi_{1}:$ [\attr{age}, \attr{overall\_status}, \attr{gender}] $\to$ [\attr{drug\_name}*]; and (ii) $\varphi_{2}:$ [\attr{overall\_status}, \attr{timeframe}, \attr{measure}] $\to$ [\attr{condition}*].  We construct attribute value generalization hierarchies (\vgh) with five levels on attributes \attr{drug name} and \attr{condition}, respectively using external ontologies (Bioprotal Medical Ontology \cite{medical_ontology}, the University of Maryland Disease Ontology \cite{disease_ontology}, and the Libraries of Ontologies from the University of Michigan \cite{ontology}). The average number of children per node in the \vgh is eight.


\eat{
\noindent \textbf{Physionet Mimic Data. } This dataset records the physiologic signals and vital signs from clinical data of patients during their hospital stay in the Boston area \cite{mimic}.  We focus on the admissions table containing 60K records with 24 attributes describing patient demographics, diagnosis, prescribed drugs, and discharge details. We define the country\_of\_birth, language and ethnicity as QI attributes.  We define two FDs: (i) $\varphi_{3}:$ [country\_of\_birth, religion*] $\to$ [language]; and (ii) $\varphi_{4}:$ [diagnosis, age] $\to$ [dose*]. The maximal level of \vgh \  on religion and language attribute are both $l=3$. 
}

\eat{
\noindent \textbf{TPC-H Data.}  We use the TPC-H dbgen data generator to generate 2M records in the LINEITEM table \cite{tpch} for scalability testing. QI attributes include linestatus, quantity and commitDate.  We define two FDs: (i) $\varphi_{5}:$ [linestatus, commitDate] $\to$ [shipDate]; and (ii) $\varphi_{6}:$ [lineNumber, quantity] $\to$ [extendedPrice]. The \vgh \ height on commitDate and extendedPrice attributes are 2 and 3, respectively. 
}

\noindent \uline{Census.} The U.S. Census Bureau provides population characteristics such as education level, years of schooling, occupation, income, and age~\cite{census}. \blue{ We select \attr{sex, age, race}, and \attr{native country} as QI attributes.}   We define two FDs: (i) $\varphi_{3}:$ [\attr{age}, \attr{education-num}] $\to$ [\attr{education*}]; and (ii) $\varphi_{4}:$ [\attr{age, industry code}, \attr{occupation}] $\to$ [\attr{wage-per-hour}*]. We construct \vgh on attributes \attr{wage-per-hour} and \attr{education} by stratifying wage and education levels according to hierarchies from US statistics~\cite{income_ontology}, and the US Department of Education~\cite{education_ontology}.  The average number of children per node is five. 

\noindent \uline{Food Inspection (Food).} This dataset contains violation citations of inspected restaurants in New York City describing the  \attr{address, borough, zipcode, violation code, violation description, inspection type, score, grade}.  \blue{We define \attr{ inspection type, borough, grade }as QI attributes.  We define two FDs: (i) $\sigma_{5}:$ [\attr{borough, zipcode}] $\to$ [\attr{address}*]; and (ii) $\sigma_{6}:$ [\attr{violation code, inspection type}] $\to$ [\attr{violation description}*].  We construct attribute \vgh on \attr{address} and \attr{violation description} by classifying streets into neighborhoods, districts, etc, and extracting topic keywords from the description and classifying the violation according to the Food Service Establishment Inspection Code \cite{food_ontology}.}  The average number of children per node in the \vgh is four.

\blue{For each dataset, we manually curate a clean instance $R_\service$ according to the defined FDs, verify with external sources and ontologies, and is used as the ground truth.  To create a dirty instance $R_\client$, we duplicate $R_\service$ to obtain $R_\client$, and use BART, an error generation benchmarking tool for data cleaning applications to inject controlled errors in the FD attributes~\cite{bart}.  We use BART to generate constraint-induced errors and random errors, and define error percentages ranging from 5\% to 25\%, with respect to the number of tuples in a table.  We use BART's default settings for all other parameters.
}

\blue{\noindent \textbf{Comparative Baseline.} }
\blue{The closest comparative baseline is PrivateClean, a framework that explores the link between differential privacy and data cleaning~\cite{krishnan}.  PrivateClean provides a mechanism to generate $\epsilon$-differentially private datasets on numerical and discrete values from which a data analyst can apply data cleaning operations.  PrivateClean proposes randomization techniques for discrete and numeric values, called Generalized Randomized Response (GRR), and applies this across all attributes.
To guarantee a privatized dataset with discrete attributes is $\epsilon$-differentially private at confidence $(1-\alpha)$, a sufficient number of distinct records is needed.  In our comparison evaluation, we set $\alpha = 0.05$, and the degree of privacy $p = 0.5$ (applied uniformly across the attributes and equivalent to $\epsilon$ for discrete attributes~\cite{krishnan}).  
Since PrivateClean does not directly propose a data cleaning algorithm, but rather data cleaning operators, we apply the well-known \attr{Greedy-Repair}~\cite{BFFR05} FD repair algorithm to generate a series of updates to correct the FD violations.  We use the \attr{transform()} operation to represent each of these updates in PrivateClean, and use the source code provided by the authors in our implementation.  We choose the \attr{Greedy-Repair} algorithm for its similarity to our repair approach; namely, to repair cells on an equivalence class basis, and to minimize a cost function that considers the number of data updates and the distance between the source and target values.
}

\blue{\noindent \textbf{Metrics.} \label{sec:privacy_exp}}
\blue{We compute the average runtime over four executions. To measure the quality of the recommended repairs, we define the \emph{repair error} of a cell as the distance between a cell's true value and its repair value.  We use the semantic distance measure, $\delta(v,v')$, in Section~\ref{sec:dist}, that quantifies the distance between two values $v$ (suppose a true value), and $v'$ (a repair value) in the attribute value generalization hierarchy, and considers the distribution of $v$ and $v'$ in the relation. \eat{Since the distance function relies on the value generalization hierarchies,}  We assume a cell's value, and its repair are both in the generalization hierarchy. The repair error of a relational instance is computed as the sum of the repair errors across all cells in the relation.  We use the absolute value of the repair error in our experiments (denoted as $\delta$).  Similar to other error metrics (such as mean squared error), lower error values are preferred.}

\noindent \textbf{Parameters.}  
\blue{Unless otherwise stated, Table~\ref{tbl:defaults} shows the range of parameter values we use, with default values in bold.}
We vary the following parameters to evaluate algorithm performance:  (i) the budget $\mc{B}$; (ii) the size of the support set $|\mathcal{S}|$ in the pricing function; (iii) $l$ (the lower bound of generalization); (i) $k$, the number of tuples in an $X$-group; and (vi) the error rate $e$ in $R_\client$.   


\subsection{Generalized Values}
\eat{Since we kept the generalized values during our cleaning process, the \client may contain general values after repair. These generalized values could have some impact on the specificity, repair accuracy and runtime. To measure the impact of these generalized values, we used the distance function to quantify the difference between the \client with generalized values and the \client containing specific values only.  \fei{What does this mean to the accuracy calc?} \yu{fixed, to make it more clear.} To measure the specificity, for each generalized value $v$, we calculate the distance $\delta (v,v')$ between $v$ and its descendant $v'$ through the entropy-based penalty distance function described in Section \ref{sec:dist}. To quantify to overall distance between ground relation and generalized relation, we normalize the distance between ground value and generalized value into four ranges [0-0.25], [0.25-0.5], [0.5-0.75], [0.75-1] to observe the impact on \client.  
}


We measure the proportion of generalized values that are returned for varying levels of the budget $\mc{B}$.  Since a repair value may be returned at any level of the \vgh, we compute the semantic distance between the generalized value $v$ and the corresponding ground value $v'$ in the \client. We normalize the distance between ($v, v'$) into four ranges:  [0-0.25], [0.25-0.5], [0.5-0.75], [0.75-1], where a distance of zero indicates $v$ and $v'$ are both ground values.

\begin{figure}[]
  \begin{minipage}[h]{.32\textwidth}
  \includegraphics[width=\textwidth]{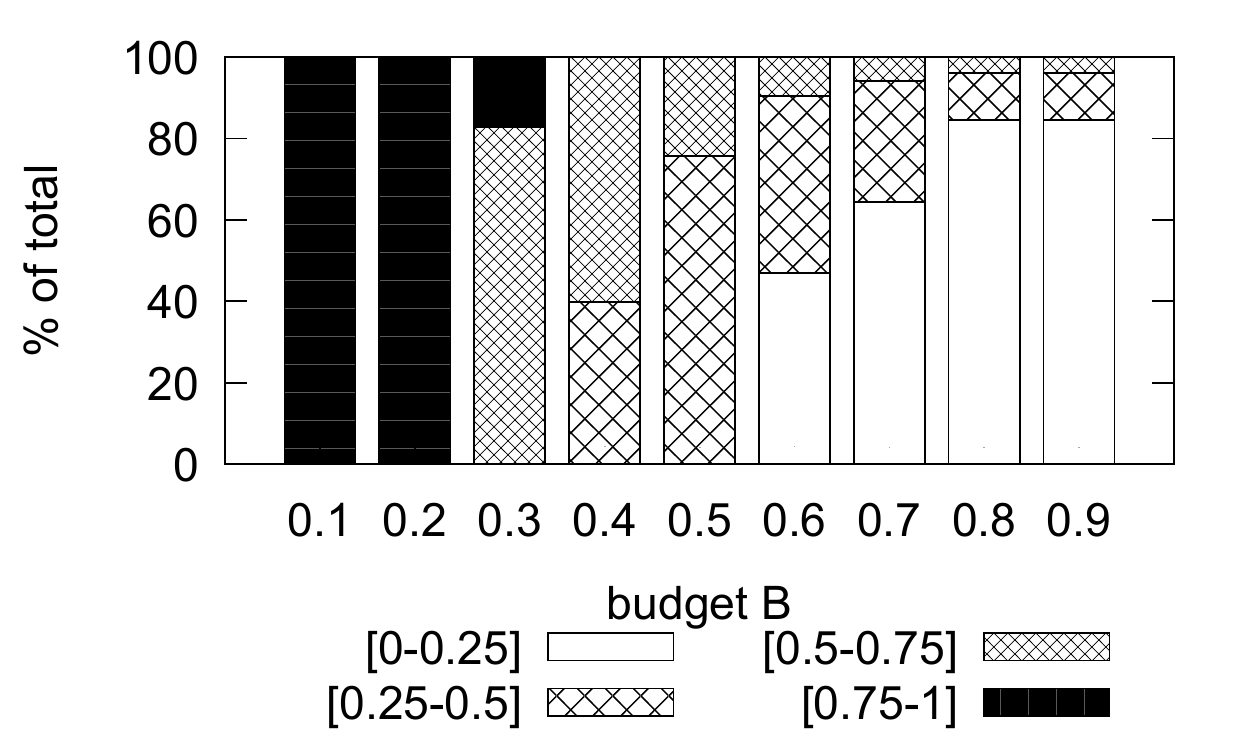}
  \captionof{figure}{Ratio of gen. values.}
  \label{fig:b_vs_clinical}
  \end{minipage}%
  \begin{minipage}[h]{.32\textwidth}
  \includegraphics[width=\textwidth]{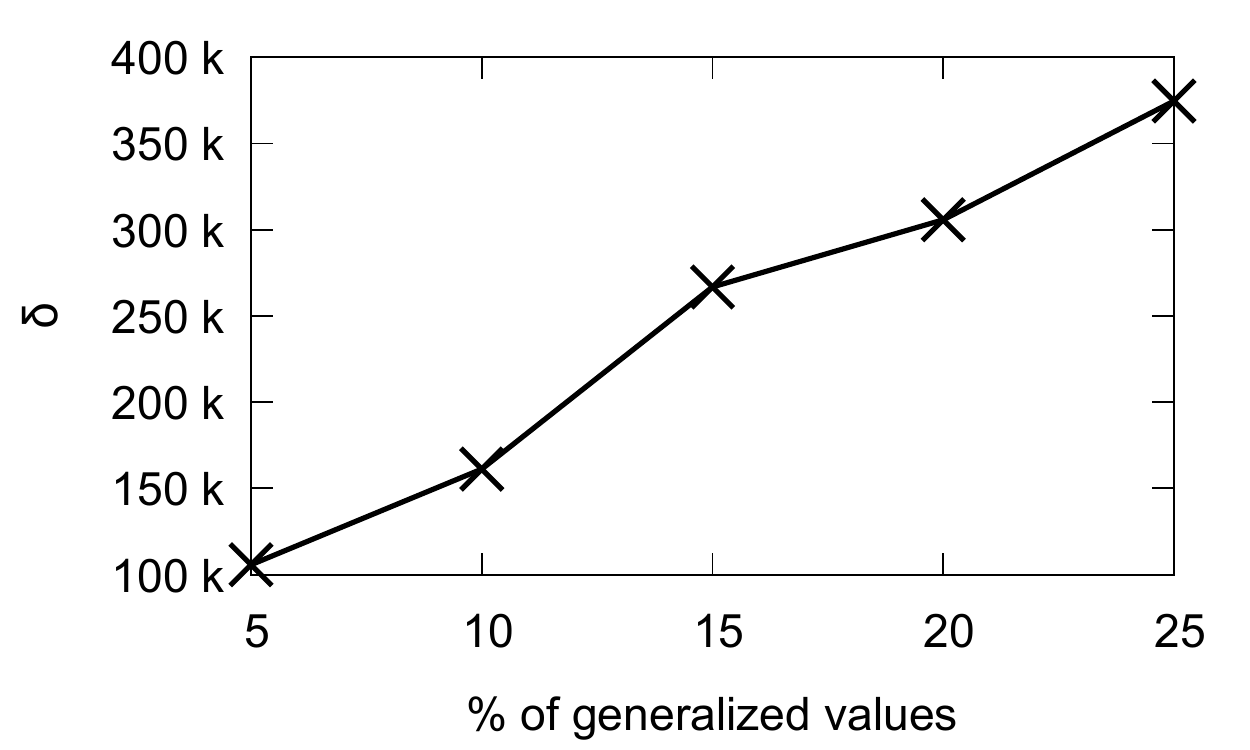}
  \captionof{figure}{Error vs. gen. values.}
  \label{fig:F1_vs_numGeneralized}
  \end{minipage}%
    \begin{minipage}[h]{.32\textwidth}
  \includegraphics[width=\textwidth]{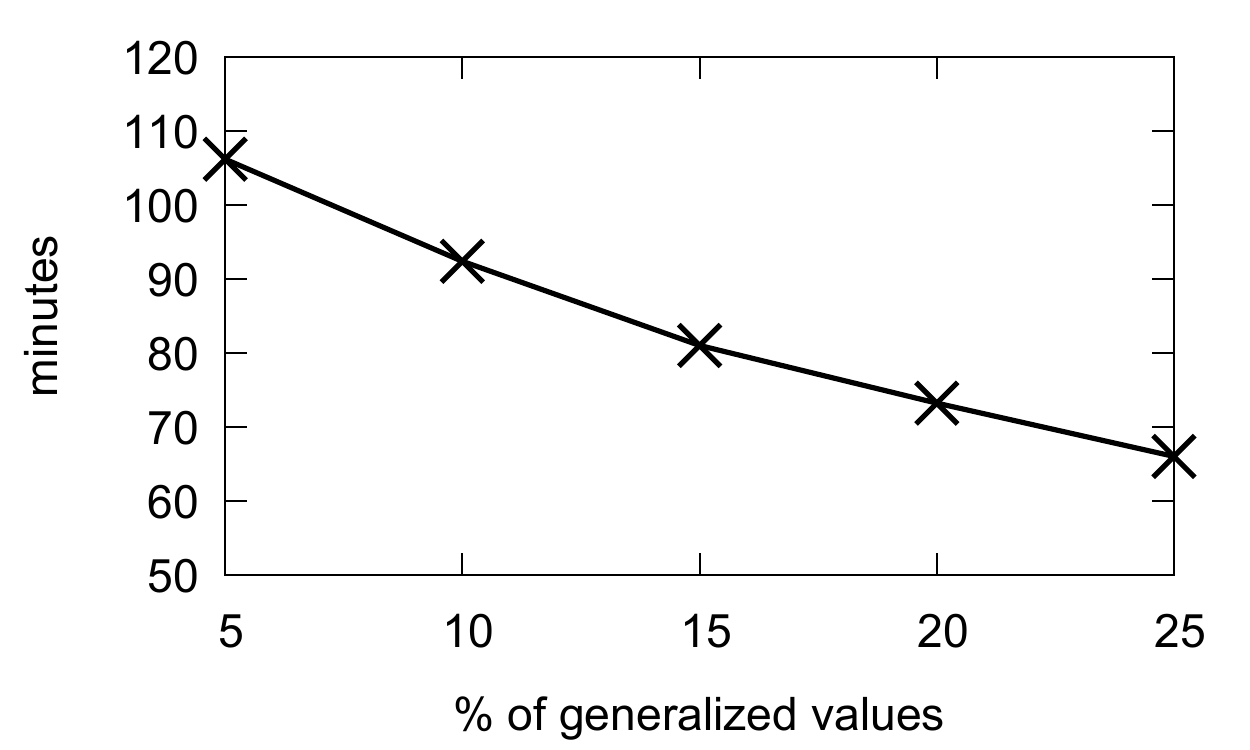}
  \captionof{figure}{Runtime vs. gen. values.}
  \label{fig:runtime_vs_numGeneralized}
  \end{minipage}%
\end{figure}

Figure \ref{fig:b_vs_clinical} shows the relative proportions for varying $\mc{B}$ values over the clinical dataset.  As expected, the results show that the proportion of generalized values at the highest levels of the \vgh occur for low $\mc{B}$ values since we can only afford (cheaper) generalized values under a constrained budget. \eat{distance reduces as we increase the budget. We observe that for the lowest $\mc{B}$ values, the errors are fixed by highly generalized repair values, therefore the distance of tuples is in range [0.75-1]. } In contrast, for $\mc{B}$ values close to 0.9, close to $85\%$ of the repair values are specific, ground values with a distance range of [0, 0.25], while the remaining $15\%$  are generalized values at the next level.  \eat{, which occur due to disclosure restrictions to satisfy our privacy requirements.}  We observe that for $\mc{B} > 0.6$ approximately 70\% of the total repair values  are very close to the ground value (where distance is at [0-0.25]), indicating higher quality repairs.
\ignore{
\begin{figure}[ht]
\begin{minipage}[h]{.32\textwidth}
  \includegraphics[width=\textwidth]{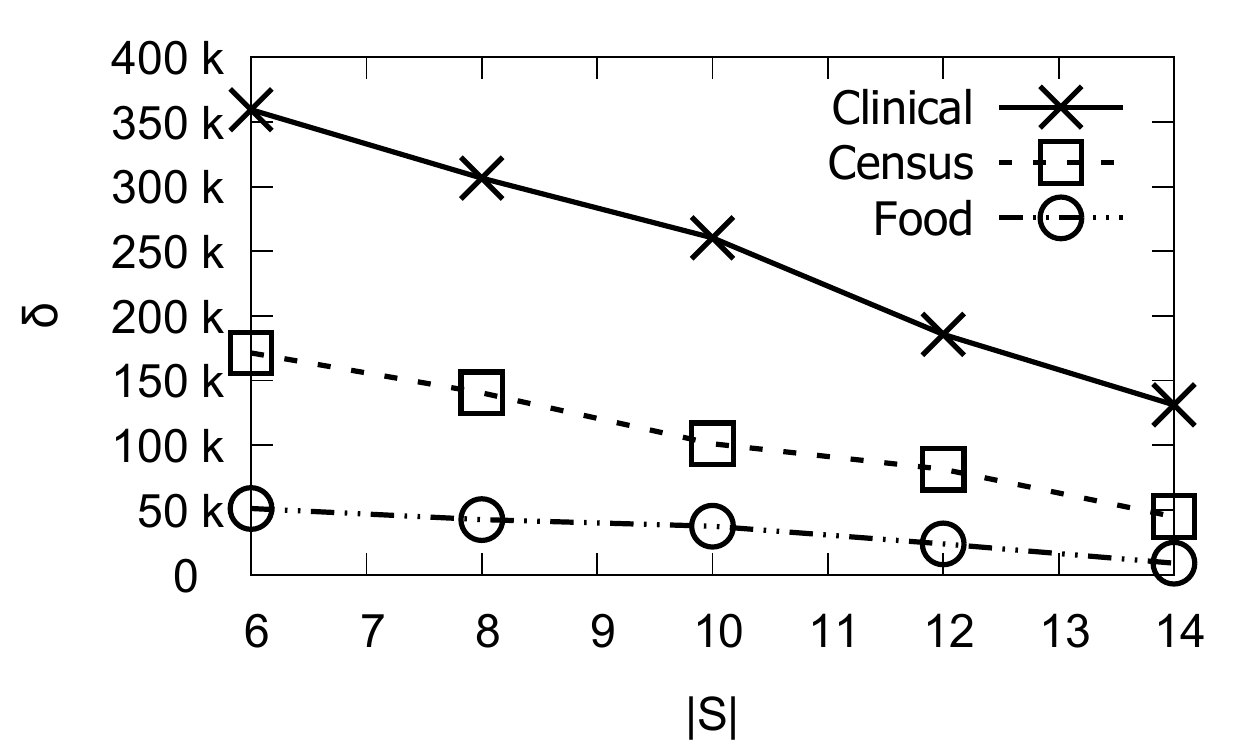}
    \captionof{figure}{Repair error vs. $|\mathcal{S}|$.}
    \label{fig:F1_vs_S}
\end{minipage}%
  \begin{minipage}[h]{.32\textwidth}
  \includegraphics[width=\textwidth]{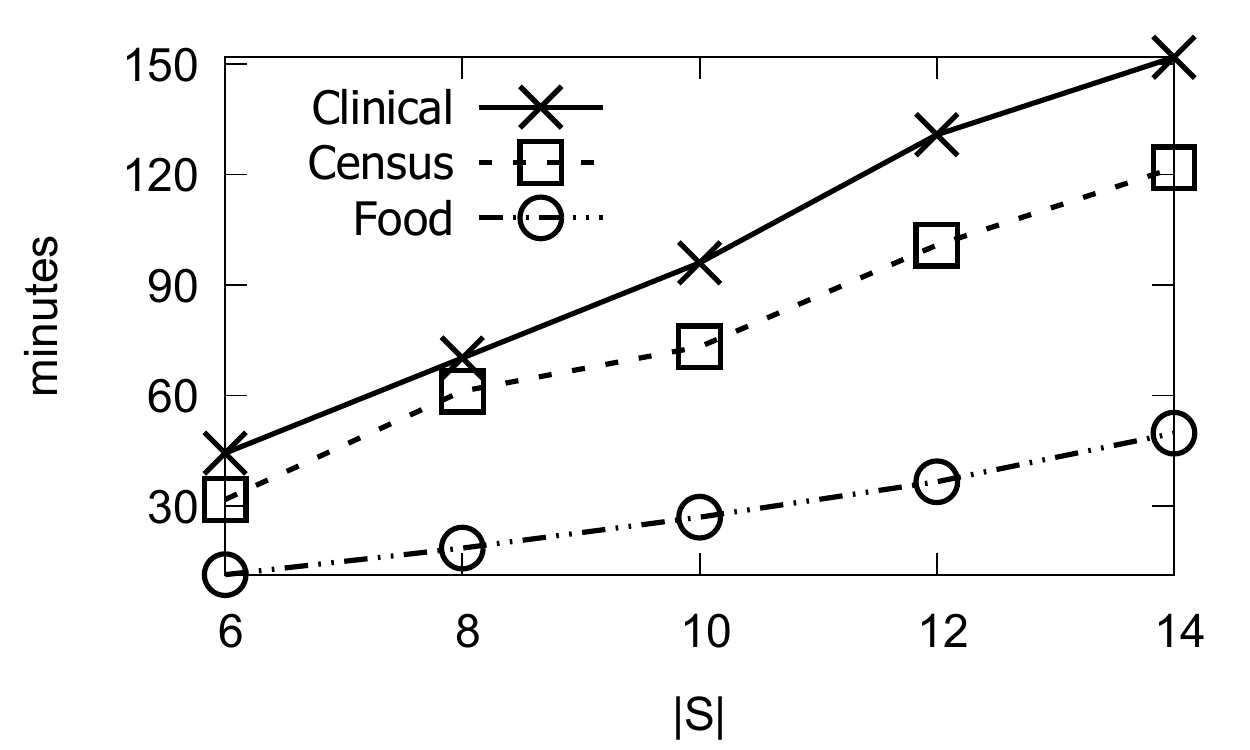}
  \captionof{figure}{Runtime  vs. $|\mathcal{S}|$.}
  \label{fig:runtime_vs_S}
\end{minipage}
  \begin{minipage}[h]{.32\textwidth}
  \includegraphics[width=\textwidth]{exp_fig/2_1_2_1.pdf}
  \captionof{figure}{Ratio of gen. values.}
  \label{fig:b_vs_clinical}
  \end{minipage}%

  \begin{minipage}[h]{.32\textwidth}
  \includegraphics[width=\textwidth]{exp_fig/F1_vs_numGeneralized.pdf}
  \captionof{figure}{Error vs. gen. values.}
  \label{fig:F1_vs_numGeneralized}
  \end{minipage}%
    \begin{minipage}[h]{.32\textwidth}
  \includegraphics[width=\textwidth]{exp_fig/runtime_vs_numGeneralized.pdf}
  \captionof{figure}{Runtime vs. gen. values.}
  \label{fig:runtime_vs_numGeneralized}
  \end{minipage}%
   \begin{minipage}[h]{.32\textwidth}
  \includegraphics[width=\textwidth]{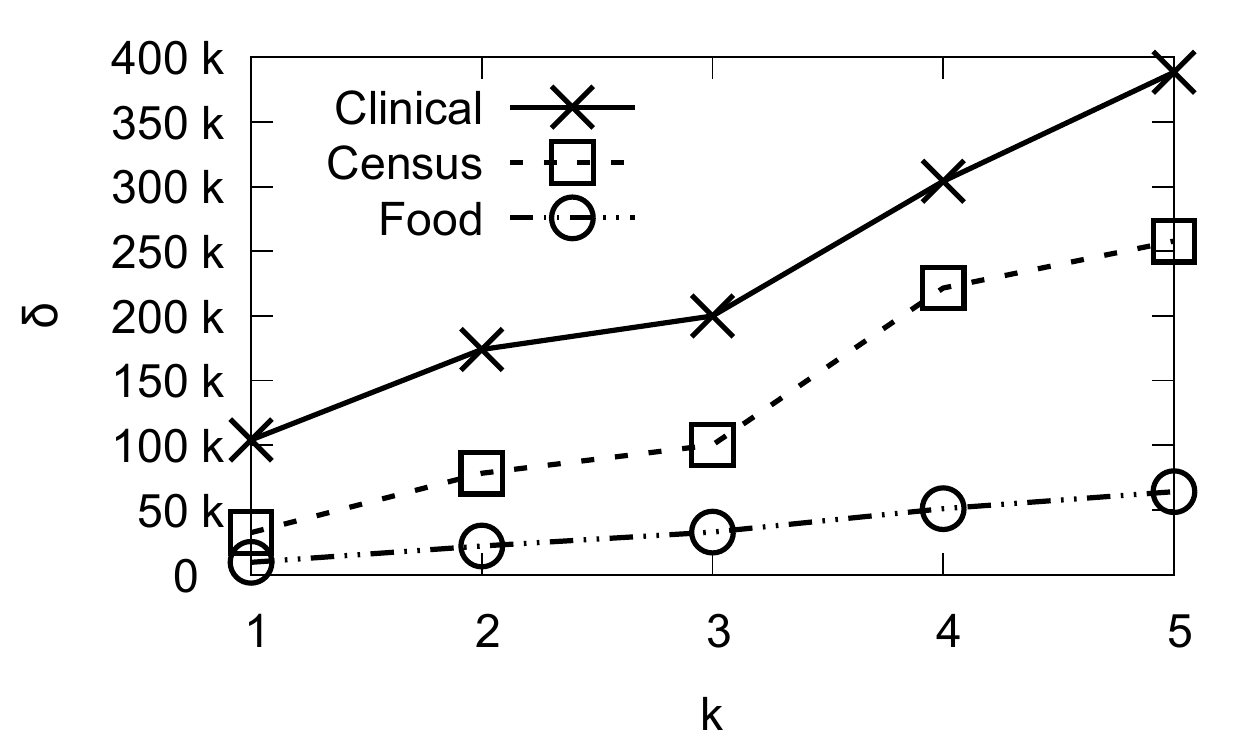}
  \captionof{figure}{Repair error vs. $k$.}
  \label{fig:F1_vs_k}
  \end{minipage}%
\end{figure}
  }

We evaluate the impact on the runtime to \blue{repair error} tradeoff when an increasing number of generalized values occur in the repaired relation.  We control the number of generalized values indirectly via the number of error cells under a constrained budget $\mc{B} = 0.1$ where it is expected that close to all repair recommendations will be general values.  Figure \ref{fig:F1_vs_numGeneralized} and Figure \ref{fig:runtime_vs_numGeneralized} show the \blue{repair error} and runtime curves over the clinical data, respectively.  As expected, we observe for an increasing number of generalized values, the \blue{repair error increases} as the constrained budget leads to more values returned at the highest * generalized value, thereby {increasing the distance between ground-truth relation and repaired relation}.  In contrast, the increased number of generalized values leads to    lower runtimes due to the increased number of unsatisfied query requests. 
\subsection{\palg Efficiency and Effectiveness}  
\eatttr{Since our pricing and privacy checking algorithm relies on the size of support set as we discussed in Section \ref{sec:pricing}. We vary the size of support size on clinical dataset to study its influence on repair accuracy and performance.  As $\mathcal{S}$ increases our accuracy increases as well since the larger support set can assign more appropriate and reasonable price to request than smaller one does. It is hard to make the original relation $R_{\service}$ indistinguishable in a small support set, so our algorithm tends to rejects requests to satisfy the privacy request, which leads to low F1 score when $\mathcal{S}$ is small. When $\mathcal{S}$ is large, the original $R_{\service}$ can be easily hidden in the large support set, and the request of privacy parameter $k$ in Algorithm \ref{alg:ppricing} is easy to reach as well, which leads to the decrease of rejected requests, and the accuracy also gets improved. However, Figure \ref{fig:runtime_vs_S} shows the runtime of support set increases as its size increases since large support set indicates the number of relations that each query needs to check for pricing also expands, which takes more execution time. We evaluate the accuracy and efficiency of our \palg data pricing algorithm  that preserves \xyl-anonymity.
}

\begin{wrapfigure}{r}{0.36\textwidth}
  \begin{center}
  \vspace{-8mm}
  \includegraphics[width=5.2cm]{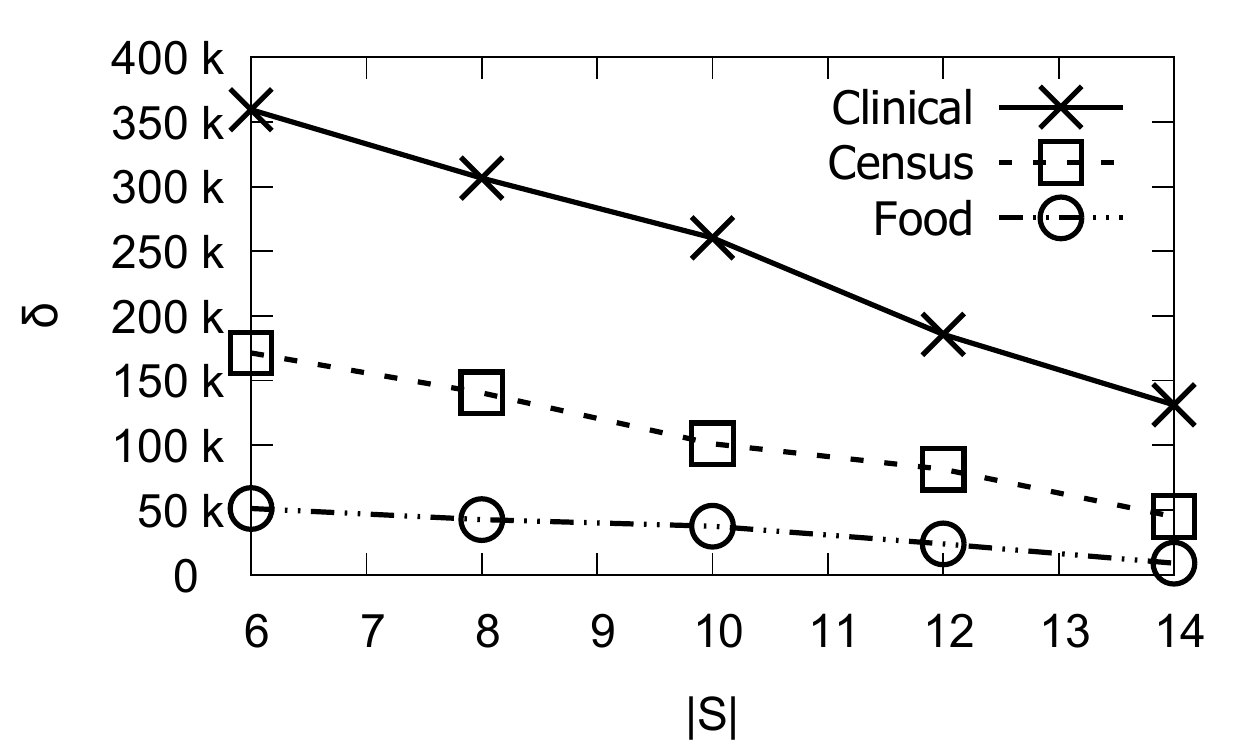}
  \vspace{-3mm}
    \captionof{figure}{Repair error vs. $|\mathcal{S}|$.}
    \label{fig:F1_vs_S}
    \vspace{-8mm}
  \end{center}
\end{wrapfigure}

The \palg algorithm relies on a support set $\mathcal{S}$ to determine query prices by summing the weights of discarded instances from the conflict set $\mathcal{C}$.  These discarded instances represent the knowledge gained by \client.  We vary the size of the initial $\mathcal{S}$ to determine its influence on the \blue{repair error ($\delta$)}, and the overall runtime.  Figure \ref{fig:F1_vs_S} shows a steady \blue{decrease} in the repair error for increasing $|\mathcal{S}|$.  As  $|\mathcal{S}|$ grows, the \service is less restrictive to answer GQs and fewer requests are declined at lower levels. \blue{As more requests are answered at these lower levels, the repair error decreases}. Figure \ref{fig:runtime_vs_S} shows that the \palg runtime  scales linearly with increasing $|\mathcal{S}|$, making it feasible to implement in practice.  From Figures \ref{fig:F1_vs_S} and \ref{fig:runtime_vs_S}, we determine that \palg achieves an average \blue{6\% reduction in the repair error at a cost of 16m runtime}.  This is expected due to the additional time needed to evaluate the larger space of instances to answer GQs. Comparing across the three datasets,  the data sizes affect runtimes as a larger number of records must be evaluated during query pricing.  This is reflected in longer runtimes for the larger clinical and census datasets.

\eat{
According to the definitoin of $E(v)$, when $v$ is ground value, the penalty $E(v)$ is 0. Moreover, our approach will only return the generalized ancestor of the origial ground value, so in our approach, the distance $\delta(v,v')$ is always based on the relation between ancestor and ground value. There is no sibling case. 
Since large support set indicates the number of relations that each query needs to check for pricing also expands, which takes more execution time. The runtimes are various on Figure \ref{fig:runtime_vs_S} because of the varies of the number of tuples on these dataset. Recall that the pricing of a query $Q$ is determined by the query result $Q(D')$ on every instance $D'$ of $|\mathcal{S}|$, as the number of tuples are different on these dataset, the query operation takes more time on large datasets (such as Clinical and Census) than small datasets (Food). }


\subsection{\palg Parameter Sensitivity}

\begin{figure}[t]
   \begin{minipage}[h]{.32\textwidth}
  \includegraphics[width=\textwidth]{7_2.pdf}
  \captionof{figure}{Runtime  vs. $|\mathcal{S}|$.}
  \label{fig:runtime_vs_S}
  \end{minipage}%
  \begin{minipage}[h]{.32\textwidth}
  \includegraphics[width=\textwidth]{5_1.pdf}
  \captionof{figure}{Repair error vs. $k$.}
  \label{fig:F1_vs_k}
  \end{minipage}%
    \begin{minipage}[h]{.32\textwidth}
  \includegraphics[width=\textwidth]{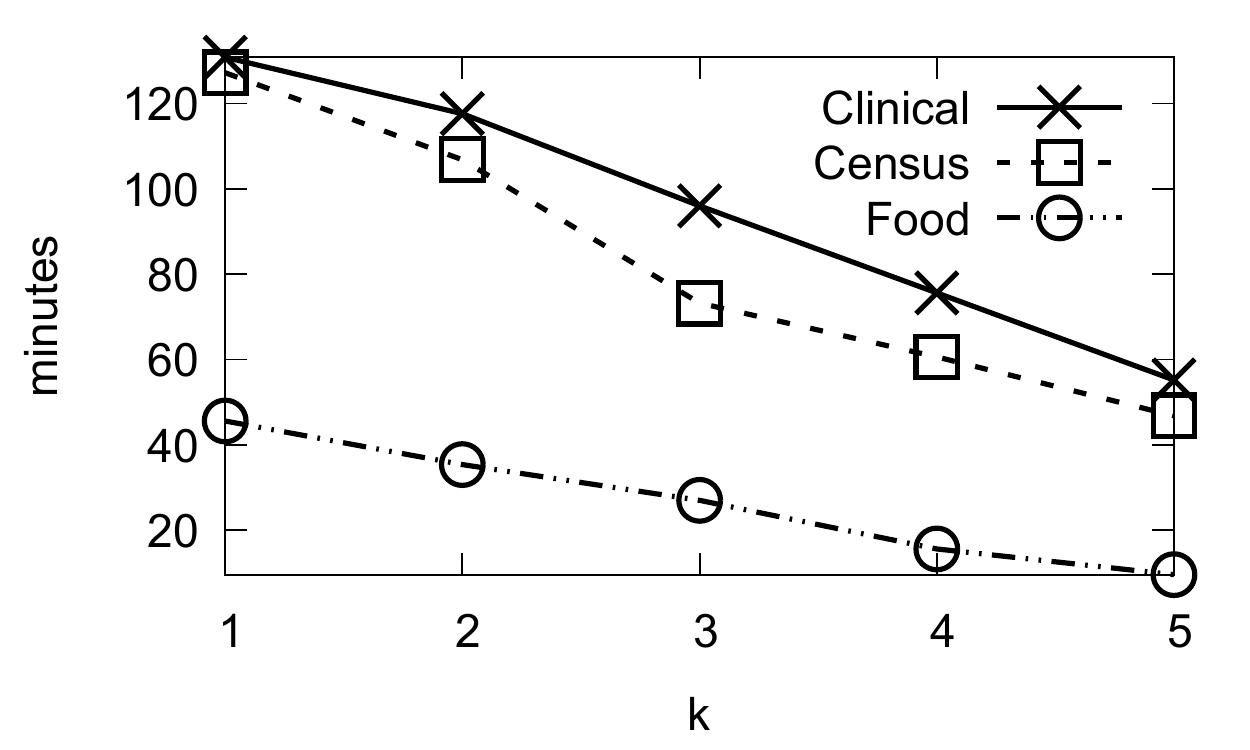}
  \captionof{figure}{Runtime vs. $k$.}
  \label{fig:runtime_vs_k}
\end{minipage}%

\begin{minipage}[h]{.32\textwidth}
  \includegraphics[width=\textwidth]{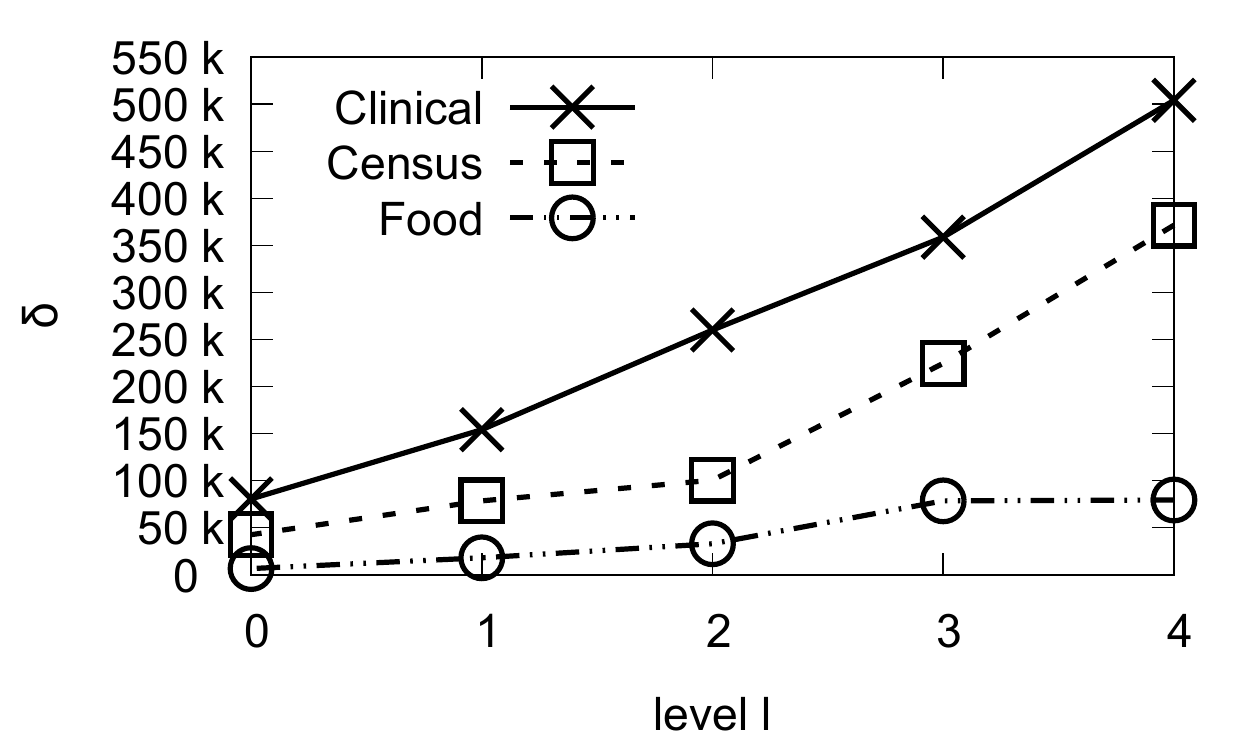}
  \captionof{figure}{Repair error vs. $l$.}
  \label{fig:F1_vs_l}
\end{minipage}%
\begin{minipage}[h]{.32\textwidth}
  \includegraphics[width=\textwidth]{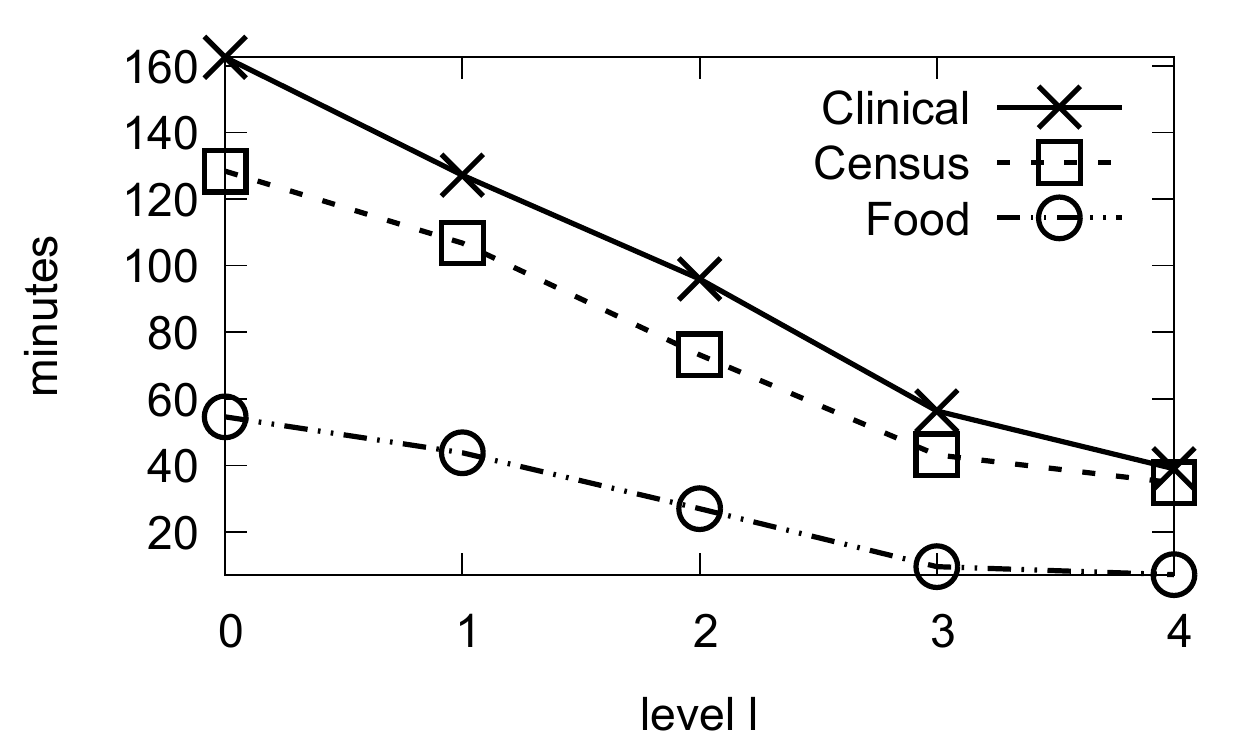}
  \captionof{figure}{Runtime vs. $l$.}
  \label{fig:runtime_vs_l}
\end{minipage}%
\begin{minipage}[h]{.32\textwidth}
  \includegraphics[width=\textwidth]{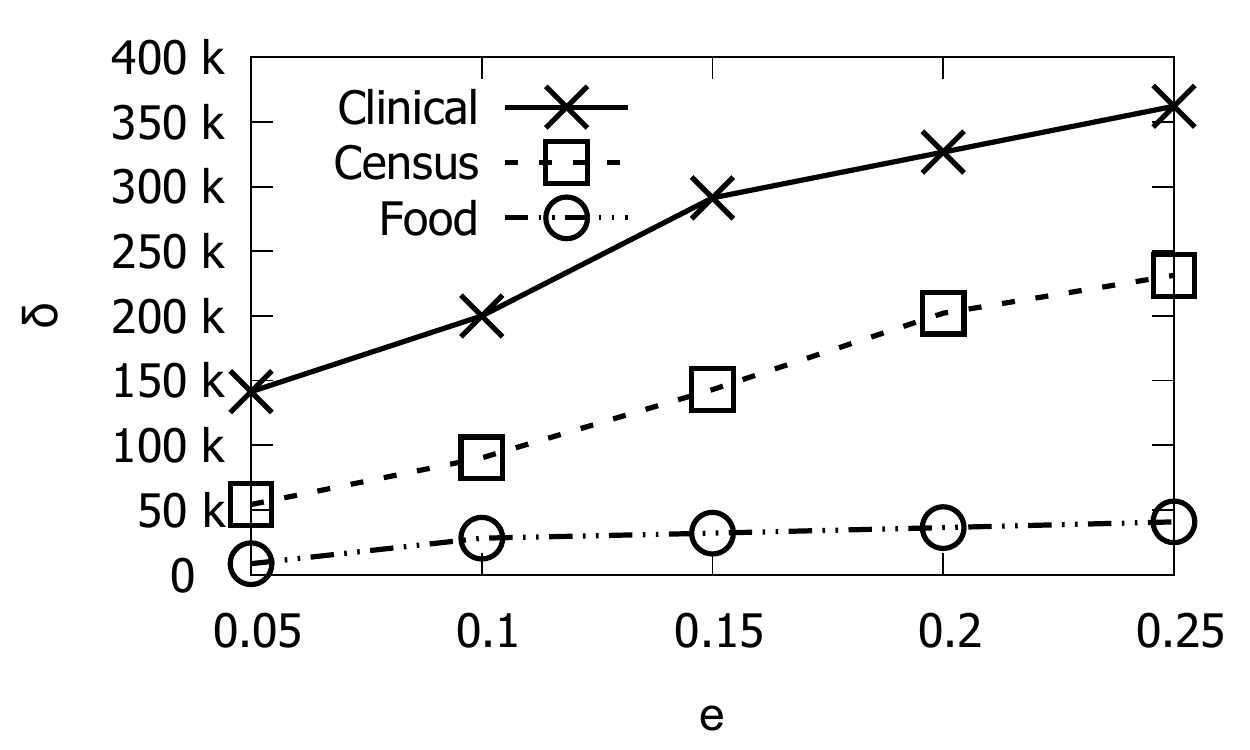}
    \captionof{figure}{Repair error vs. $e$. }
 \label{fig:F1_vs_e_3datasets}
\end{minipage}%

\begin{minipage}[h]{.32\textwidth}
  \includegraphics[width=\textwidth]{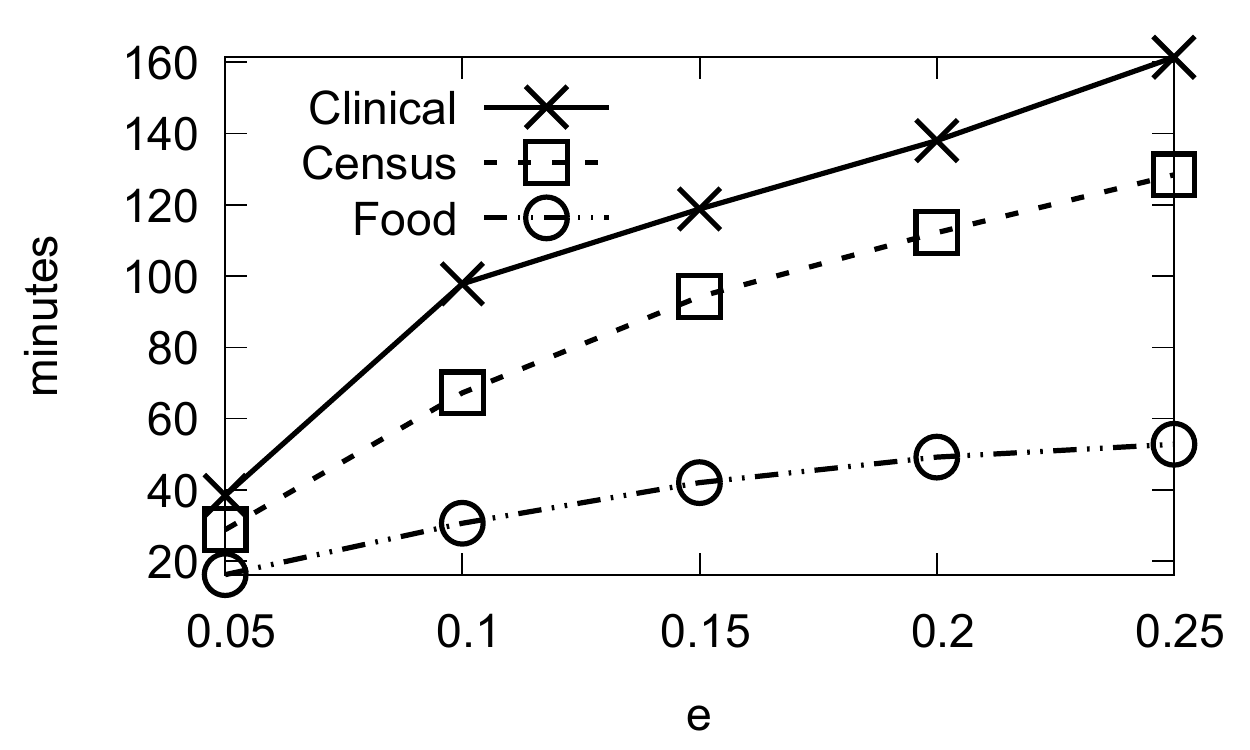}
    \captionof{figure}{Runtime vs. $e$. }
 \label{fig:runtime_vs_e_3datasets}
\end{minipage}%
\begin{minipage}[h]{.32\textwidth}
  \includegraphics[width=\textwidth]{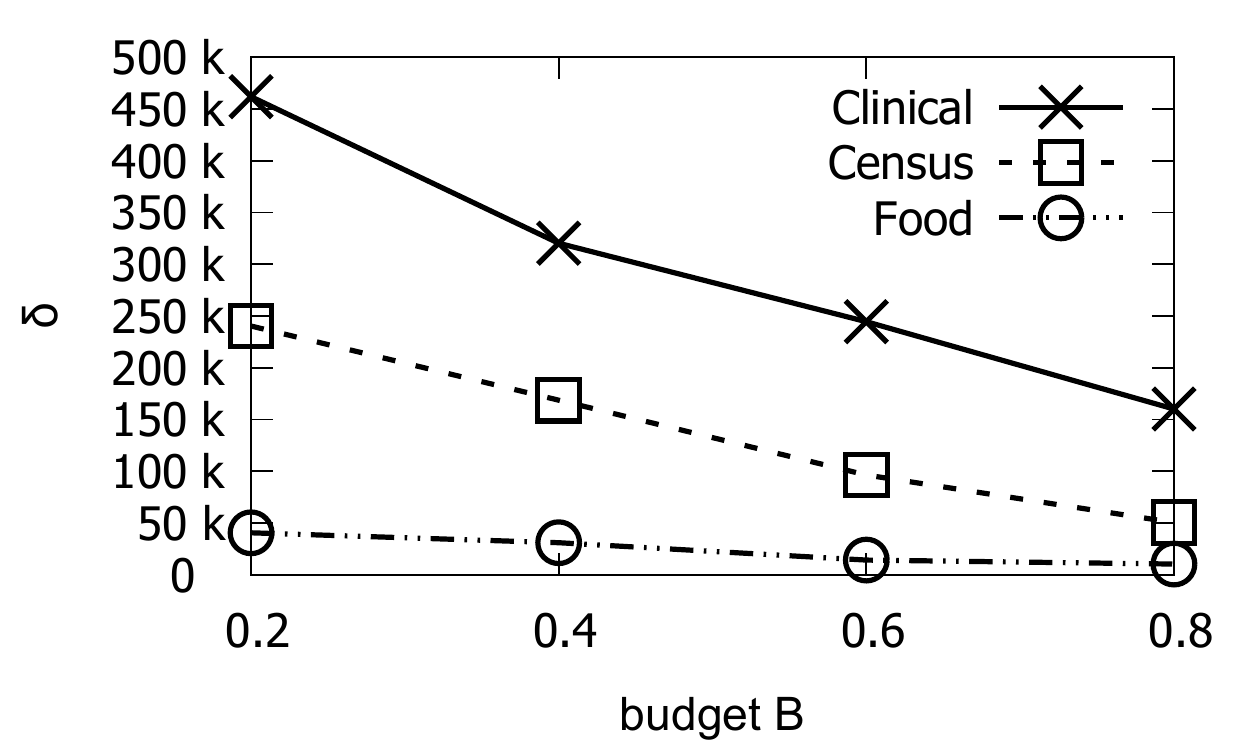}
    \captionof{figure}{Repair error vs. $\mc{B}$. }
 \label{fig:F1_vs_b_3datasets}
\end{minipage}%
\begin{minipage}[h]{.32\textwidth}
  \includegraphics[width=\textwidth]{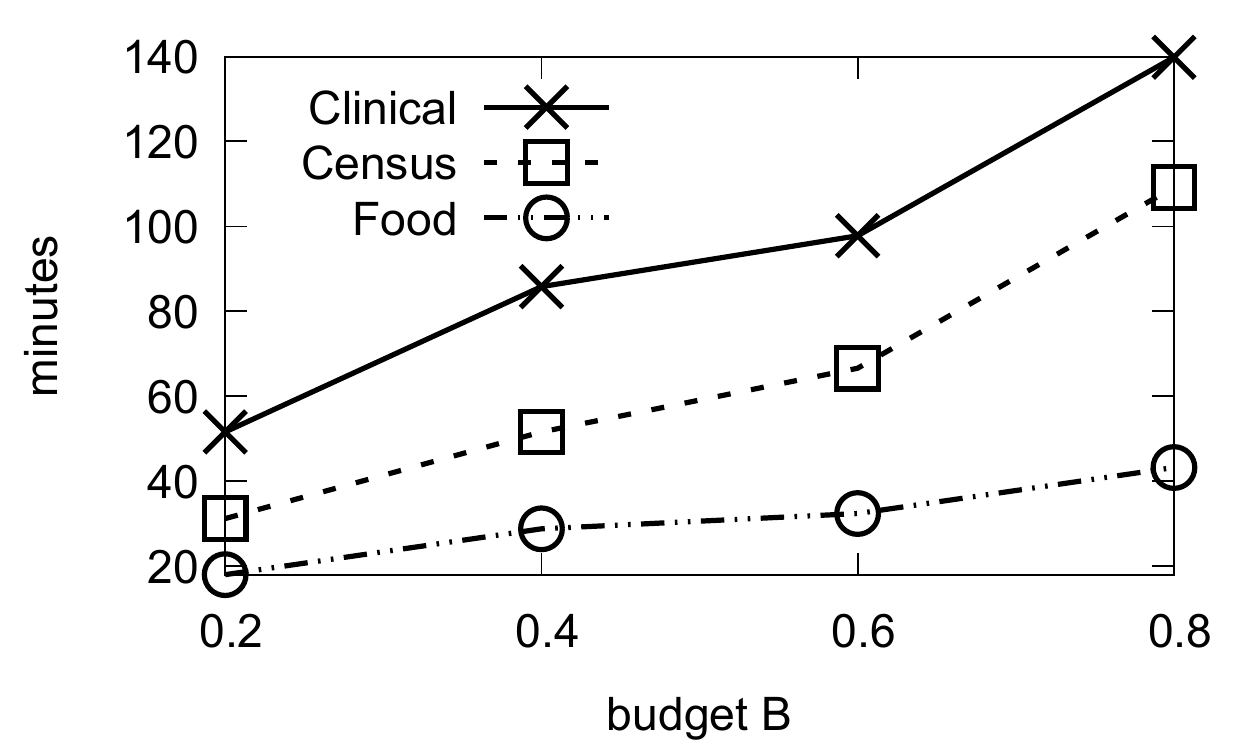}
    \captionof{figure}{Runtime vs. $\mc{B}$. }
 \label{fig:runtime_vs_b_3datasets}
\end{minipage}%

\end{figure}

We vary the parameters $k$, GQ level $l$, error rate $e$, budget $\mc{B}$, and measure their influence on \calg \ \blue{repair error} and runtime over all three datasets. We expect that enforcing more stringent privacy requirements through larger $k$ and $l$ values will result in \blue{larger repair errors}.  \blue{Figures \ref{fig:F1_vs_k}} to \ref{fig:runtime_vs_l} do indeed reflect this intuition.  

In Figure \ref{fig:F1_vs_k}, \calg experiences larger repair errors for increasing $k$ as generalizations to conceal sensitive values become increasingly dependent on the attribute domain and its VGH  i.e., \xyl-anonymity indicates there must be at least $k$ values at a given level $l$ in the VGH.  Otherwise, the data request is denied. Figure~\ref{fig:runtime_vs_k} shows that for increasing $k$, runtimes decrease linearly as query requests to satisfy more stringent $k$ become more difficult.  On average, we observe that \blue{an approximate 10\% improvement in runtime leads to a 7\% increase in the repair error} for each increment of $k$.  Figures \ref{fig:F1_vs_l} and \ref{fig:runtime_vs_l} show the \blue{repair error} and runtime, respectively, as we vary the query level parameter $l$.  The \blue{repair error increases}, particularly after $l = 3$ as more stringent privacy requirements are enforced, i.e., $l$ distinct values are required at each generalization level.  This makes satisfying query requests more difficult, leading to unrepaired values and lower runtimes, as shown in Figure \ref{fig:runtime_vs_l}.

Figure \ref{fig:F1_vs_e_3datasets} shows the \blue{repair error $\delta$ increases} as we scale the error rate $e$. For a fixed budget, increasing the number of FD errors leads to a decreasing budget for each FD error. This makes some repairs unaffordable for the \client, leading to unrepaired values and \blue{an increased number of generalized repair values}. This situation can be mitigated if we increase the budget $\mc{B}$. \blue{As expected, Figure \ref{fig:runtime_vs_e_3datasets} shows that the \calg \ runtime increases for an increasing number of FD violations, due to the larger overhead to compute more equivalance classes, compute prices to answer queries, and to check consistency in the \client. }

\blue{Figures \ref{fig:F1_vs_b_3datasets} and \ref{fig:runtime_vs_b_3datasets} show the repair error and runtime, respectively, as we vary the budget $\mc{B}$.  For increasing budget allocations, we expect the \service to recommend more ground (repair) values, and lower repair error  values, as shown in Figure~\ref{fig:F1_vs_b_3datasets}.   Given the larger budget allocations, the \service is able to answer a larger number of query requests, and must compute their query prices, thereby increasing algorithm runtime. We observe that an average 14\% reduction in the repair error leads to an approximate 7\% increase in runtime.} 
\subsection{Comparative Evaluation}\label{sec:comparative}
\eatttr{We compare our approach against PrivateClean \cite{krishnan}. PrivateClean applies another privacy model--differential privacy to the data cleaning process. In the PrivateClean paper, it does not introduce any cleaning algorithm, therefore, we takes the existing cardinality minimal data cleaning approach, which transforms a minimal number of data values in the dirty database to satisfy FDs $\Sigma$ \cite{BFFR05}. In our tests, we set the PrivateClean parameter $p = 0.5$ according to its recommendation in \cite{krishnan}. 
The reason is that PrivateClean approach protects the privacy via differential privacy by replacing attribute values with random domain values, which not only affects the finding matched record process, but also makes it hard to repair the error back to original value. Although PrivateClean provides strong guarantees to control the distortion in query answering, the utility of the data is severely reduced for data cleaning purposes. As acknowledged by the authors, identifying and reasoning about errors over randomized data is hard.   This makes it difficult to implement in practice.  We envision that our approach \calg is a compromise towards providing a \xyl-anonymous privacy solution while still preserving good levels of data utility. 
}
\eat{I checked the SIGMOD 2019 paper by Ihab, where they used differential privacy on the matching process of entity resolution. Even in their paper, they argued that PrivateClean is in different setting, where no cleaning algorithm is provided and the cleaning is relied on data analyst. So in their paper, they didn't compare against PrivateClean in their experiment. From what I read and the related work and references of that paper, currently only two similar papers are working on Privacy + Cleaning. One is PrivateClean, another one is Dhruv's paper. For fair comparison, we can use the privacy algorithm of PrivateClean, which is differential privacy, then use minimal changes for cleaning since only one dataset involved. So the comparison is differential privacy + minimal changes vs SafeClean. We can argue that although DP provides better privacy preservation, but the cost is the repair accuracy. }

\blue{We compare \calg to PrivateClean, a framework for data cleaning on locally differentially private relations~\cite{krishnan}.  Section~\ref{sec:expsetup} describes the baseline algorithm parameter settings and configuration.  Despite the differing privacy models, our evaluation aims to explore the influence of increasing error rates on the repair error $\delta$, and algorithm runtimes using the Clinical dataset.  We hope these results are useful for practitioners to understand qualitative and performance trade-offs between the two privacy models.  For \calg, we measure total time of the end-to-end process from error detection to applying as many repairs as the budget allows.  In PrivateClean, we measure the time to privatize the $R_\client$, error detection, running the \attr{Greedy-Repair} FD repair algorithm~\cite{BFFR05}, and applying the updates via the \attr{Transform} operation. }
\begin{wrapfigure}{r}{0.36\textwidth}
  \begin{center}
  \vspace{-8mm}
  \includegraphics[width=5.2cm]{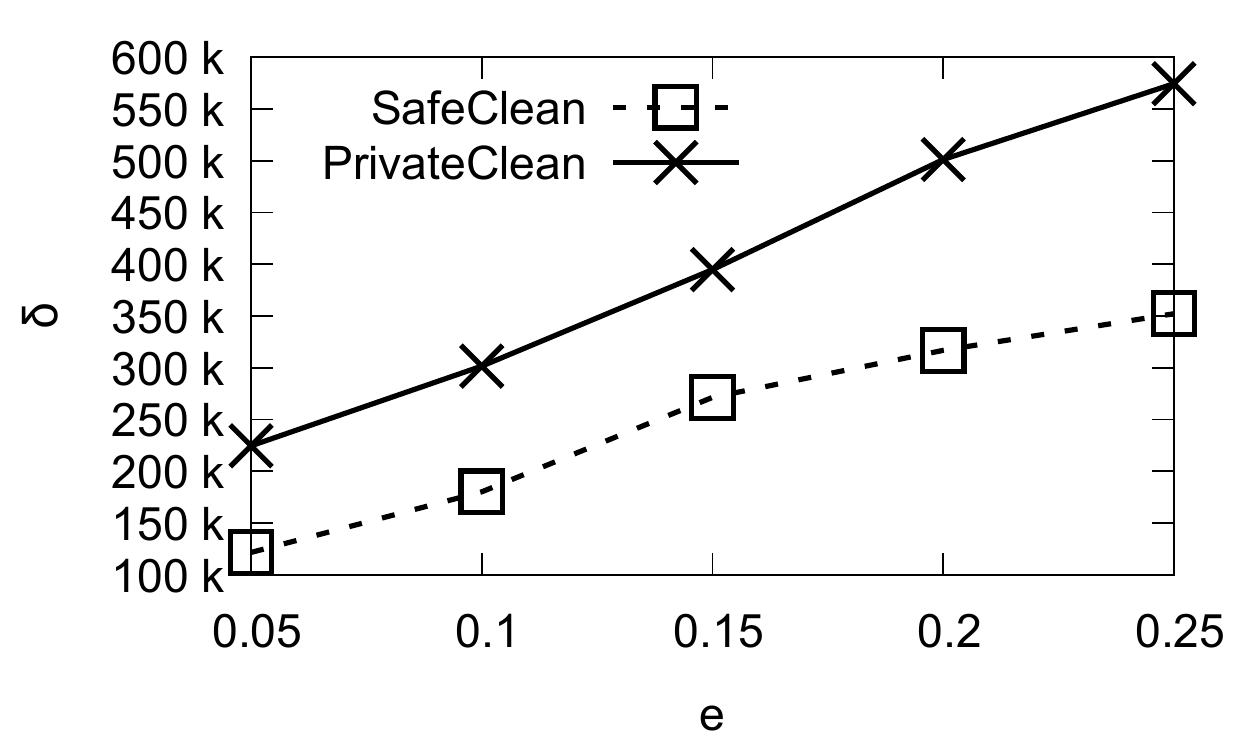}
     \vspace{-3mm}
    \captionof{figure}{Comparative repair error.}
 \label{fig:F1_vs_e_privateClean_xyl}

  \includegraphics[width=5.2cm]{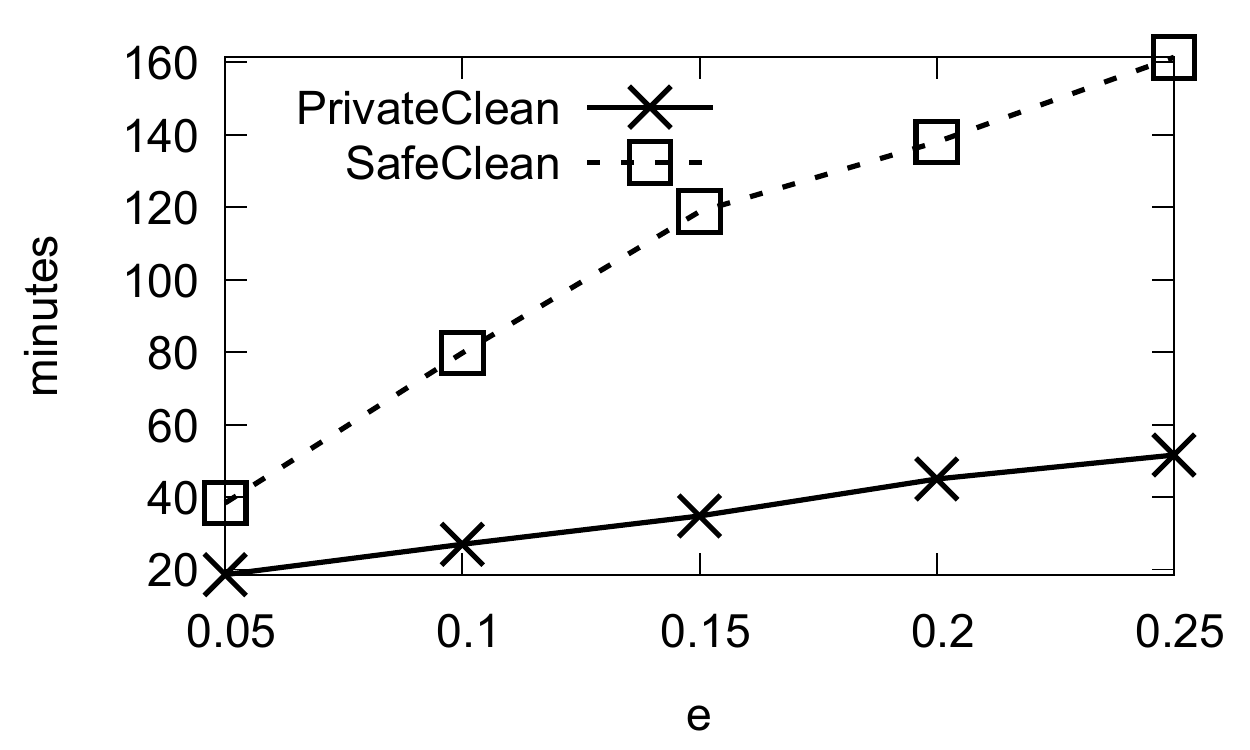}
  \vspace{-4mm}
 \captionof{figure}{Comparative runtime.} 
 \label{fig:runtime_vs_e_privateClean_xyl}
  \end{center}
  \vspace{-6mm}
\end{wrapfigure}
\blue{For PrivateClean, we measure the repair error $\delta(v,v')$ for source value $v$ and target (clean) value $v'$, as recommended by \attr{Greedy-Repair}, where both $v, v'$ are ground values.}

Figure \ref{fig:F1_vs_e_privateClean_xyl} shows the comparative \blue{repair error} between  \calg and PrivateClean as we vary the error rate $e$. \calg achieves an average \blue{$-41\%$} lower repair error than PrivateClean. This poor performance by PrivateClean is explained by the underlying data randomization used in differential privacy, which provides strong privacy guarantees, but poor data utility, especially in data cleaning applications. As acknowledged by the authors, identifying and reasoning about errors over randomized response data is hard~\cite{krishnan}.  \blue{This randomization may update rare values to be more frequent, and similarly, common values to be more rare.  This leads to more uniform distributions (where stronger privacy guarantees can be provided), but negatively impact error detection and cardinality-minimal data repair techniques that rely on  attribute value frequencies to determine repair values~\cite{BFFR05}.}  We envision that \calg is a compromise towards providing an \xyl-anonymous instance while preserving data utility. 

\eat{It sets a probability $p$ ($p=0.5$ as they recommended in their paper) to randomly replace attribute values with some other values from the same domain, which dramatically increases the $\delta$.}  \calg's \blue{lower repair error} comes at a runtime cost, as shown in Figure \ref{fig:runtime_vs_e_privateClean_xyl}. As we scale the error rate,  \calg's runtime scales linearly due to the increased overhead of data pricing.  Recall the pricing mechanism in Section \ref{sec:gprice}, the price of query $Q$ is determined by the query answer over the relation $D$ and its neighbors $D'$ in the support set $\mc{S}$. \blue{In contrast, PrivateClean does not incur such an overhead due to its inherent data randomization.  There are opportunities to explore optimizations to lower \calg's overhead to answer query requests and compute prices to answer each query.  In practice, optimizations can be applied to improve overall performance, including decreasing the parameter $k$ according to application requirements, and considering distributed (parallel) executions of query processing over partitions of the data.  \calg aims to provide a balanced approach towards achieving data utility for data cleaning applications while protecting sensitive values via  data pricing and \xyl-anonymity. }

\section{Related Work} 
\label{sec:related}

\noindent \textbf{Data Privacy and Data Cleaning. }
We extend the PACAS framework introduced in~\cite{huang2018pacas}, which focuses on repairs involving ground values.  \blue{While the prior PACAS framework prices and returns generalized values, the \client does not allow generalized values in its relation $R_\client$.  Instead, the \client instantiates a grounding process to ground any generalized values returned by the \service.  In this work, we remove this limitation by re-defining the notion of consistency between $R_\client$ and the set of FDs $\Sigma$ to include generalized values. We have also extended the budget allocation scheme to consider allocations according to the proportion of errors in which cells (in an equivalence class) participate.  This improves the PACAS framework to be more adaptive to the changing number of unresolved errors instead of only assigning fixed allocations.}   Our revised experiments, \blue{using repair error $\delta$ as a measure of accuracy}, show the influence of generalized repairs along varying parameters towards improved efficiency and \blue{effectiveness}.



\blue{There has been limited work that considers data privacy requirements in data cleaning.  Jaganathan and Wright propose a privacy-preserving protocol between two parties that imputes missing data using a lazy decision-tree imputation algorithm \cite{jagannathan}.   Information-theoretic metrics have been used to quantify the information loss incurred by disclosing sensitive data values~\cite{huangparc,CG18}.  As mentioned previously, PrivateClean provides a framework for data cleaning on local deferentially private relations using a set of generic user-defined cleaning operations showing the trade-off between privacy bounds and query accuracy~\cite{krishnan}.  While PrivateClean provides stronger privacy bounds than PACAS, these bounds are only applicable under fixed query budgets that limit the number of allowed queries (only aggregation queries), which is difficult to enforce in interactive settings. Our evaluation has shown the significant repair error that PrivateClean incurs due to the inherent randomization needed in local differential privacy.}

\eat{Furthermore, the use of differential privacy in practical settings has been limited to special use cases requiring large data samples.  In data cleaning settings, where specific values are often required, the randomization process in differential privacy leads to low utility and untrustworthy values.
PrivateClean is a framework for data cleaning and approximate query processing on locally differentially private relations that combines data cleaning and local differential privacy~\cite{krishnan}. Based on generalized randomized response, PrivateClean presents user-defined data cleaning operations in the form of extraction, merging, and transformation. While generic user-defined operations are given, PrivateClean relies on the user to provide the specific cleaning semantics. In addition, as shown in our evaluation, error detection and cleaning over randomized data is difficult, and decreases data utility.}

\blue{Ge et. al., introduce APEx, an accuracy-aware, data privacy engine for sensitive data exploration that allow users to pose adaptive queries with a required accuracy bound.  APEx returns query answers satisfying the bound and guarantees that the entire data exploration process is differentially private with minimal privacy loss.  Similar to PACAS, APEx allows users to interact with private data through declarative queries, albeit specific aggregate queries (PACAS has no such restriction).  APEx enable users to perform specific tasks over the privatized data, such as entity resolution.  While similar in spirit, APEx uses the given query accuracy bound to tune differential privacy mechanisms to find a privacy level $\epsilon$ that satisfies the accuracy bound.  In contrast, PACAS applies generalization to target predicates in the given query requests, without randomizing the entire dataset, albeit with looser privacy guarantees, but with higher data utility.
}

\noindent \textbf{Dependency Based Cleaning. } Declarative approaches to data cleaning (cf.~\cite{bertossi13} for a survey) have focused on achieving consistency \blue{against} a set of dependencies such as FDs and inclusion dependencies and their extensions. Data quality semantics are declaratively specified with the dependencies, and data values that violate the dependencies are identified as errors, and repaired~\cite{fan09,chiang11,chiang14,kolahi,ilyas2015trends,BKCGS17}. There are a wide range of cleaning algorithms, based on user/expert feedback~\cite{gokhale,wang12,yakout}, master database~\cite{fan12,fan09}, knowledge bases or crowdsourcing~\cite{chu}, probabilitic inference \cite{rekatsinas2017holoclean, yu2019piclean}.  Our repair algorithm builds upon these existing techniques to include data disclosure requirements of sensitive data and suggesting generalized values as repair candidates.

\noindent \textbf{Data Privacy. }
PPDP uses generalization and suppression to limit data disclosure of sensitive values \cite{sweeney,samarati}.  The generalization problem has been shown to be intractable\ignore{~\cite{meyerson}}, where optimization\ignore{\cite{iyengar,sweeney,samarati}}, and approximation\ignore{~\cite{aggarwal,bayardo}} algorithms have been proposed (cf.~\cite{fung} for a survey). Extensions have proposed tighter restrictions to the baseline $k$-anonymity model to protect against similarity and skewness attacks by considering the distribution of the sensitive attributes in the overall population in the table~\cite{fung}.  Our extensions to \xy-anonymity to include semantics via the \vgh \ can be applied to existing PPDP techniques to semantically enrich the generalization process such that semantically equivalent values are not inadvertently disclosed. 

In differential privacy, the removal, addition or replacement of a single record in a database should not significantly impact the outcome of any statistical analysis over the database~\cite{dwork}.  An underlying assumption requires that the service provider know in advance the set of queries over the released data. This assumption does not hold for the interactive, service-based data cleaning setting considered in this paper.  \blue{PACAS aims to address this limitation by marrying PPDP and data pricing in interactive settings.}

\noindent \textbf{Data Pricing. }
We use the framework by Deep and Koutris that provides a scalable,  arbitrage-free pricing for SQL queries over relational databases  \cite{deep}.  Recent work has considered pricing functions to include additional properties such as being {\em reasonable} (differentiated query pricing based on differentiated answers), and {\em predictable}, {\em non-disclosive} (the inability to infer unpaid query answers from query prices) and {\em regret-free} (asking a sequence of queries during multiple interactions is no more costly than asking at once)~\cite{balazinska}. We are investigating  extensions of \palg to include some of these properties.


\section{Conclusions} \label{sec:conc}

We present PACAS, a data cleaning-as-a-service framework that preserves \xyl-anonymity in a service provider with sensitive data, while resolving errors in a client data instance.  \eat{We introduced generalized databases and queries, along with an extended data pricing scheme that allows the target to securely query the master.  We presented a data repair algorithm that resolves inconsistencies in the target and uses the notion of generalized repair values to protect the privacy of the master while improving consistency in the target. We proposed \palg, a data pricing algorithm that enforces \xyl-anonymity in our framework.}
\blue{PACAS anonymizes sensitive data values by implementing a data pricing scheme that assigns prices to requested data values while satisfying a given budget.  We propose generalized repair values as a mechanism to obfuscate sensitive data values, and present a new definition of consistency with respect to functional dependencies over a relation.  To adapt to the changing number of errors in the database during data cleaning, we propose a new budget allocation scheme that adjusts to the current number of unresolved errors.  We believe that PACAS provides a new approach to privacy-aware cleaning} that protects sensitive data while offering high data cleaning utility, as data markets become increasingly popular.  As next steps, we are investigating: (i)  optimizations to improve the performance of the data pricing modules; and (ii) extensions to include price negotiation among multiple service providers and clients.




\end{document}